\documentclass[oneside,english]{amsart}
\usepackage[T1]{fontenc}
\usepackage[latin9]{inputenc}
\usepackage{babel}
\usepackage{mathtools}
\usepackage{amstext}
\usepackage{amsthm}
\usepackage{amssymb}
\usepackage{graphicx}
\usepackage[authoryear]{natbib}
\usepackage[unicode=true]
 {hyperref}

\makeatletter

\newcommand{\lyxdot}{.}

\numberwithin{equation}{section}
\numberwithin{figure}{section}
\theoremstyle{plain}
\newtheorem{thm}{\protect\theoremname}
\theoremstyle{definition}
\newtheorem{defn}[thm]{\protect\definitionname}

\makeatother

\providecommand{\definitionname}{Definition}
\providecommand{\theoremname}{Theorem}

\begin{document}
\title[Candidate Incentive Distributions]{Candidate Incentive Distributions: How voting methods shape electoral
incentives}
\author{Marcus Ogren}
\begin{abstract}
We evaluate the tendency for different voting methods to promote political
compromise and reduce tensions in a society by using computer simulations
to determine which voters candidates are incentivized to appeal to.
We find that Instant Runoff Voting incentivizes candidates to appeal
to a wider range of voters than Plurality Voting, but that it leaves
candidates far more strongly incentivized to appeal to their base
than to voters in opposing factions. In contrast, we find that Condorcet
methods and STAR (Score Then Automatic Runoff) Voting provide the
most balanced incentives; these differences between voting methods
become more pronounced with more candidates are in the race and less
pronounced in the presence of strategic voting. We find that the incentives
provided by Single Transferable Vote to appeal to opposing voters
are negligible, but that a tweak to the tabulation algorithm makes
them substantial.
\end{abstract}

\maketitle

\section{Introduction\label{sec:Introduction} }

What voting methods would be the most effective for reducing a society\textquoteright s
internal tensions? Often, politicians and elected officials are faced
with choices that influence how divided their society will be. Secure
government benefits for their own group at the expense of others,
or pursue the common good? Attack rival factions, or take a conciliatory
approach? When politicians compete to outbid one another on ethnic
or divisive partisan issues, we expect polarization and the risk of
a civil war to increase; when they opt for compromise and use inclusive
rhetoric, we expect it to promote stability.

This suggests that a society should be able to improve its long-term
stability by using voting methods that give politicians an electoral
incentive to appeal to a wide swath of the electorate; were politicians
incentivized to appeal to all voters equally, the zero-sum politics
of polarization and ethnic conflict would be a much weaker campaign
strategy. Such observations form the basis of the theory of centripetalism,
which seeks to counter the centrifugal forces of ethnic outbidding
with centripetal incentives to attract support from ethnicities and
political factions other than one\textquoteright s own.\footnote{For a greater discussion of centripetalism, see \citet{horowitzoriginal}
and \citet{reilly_divided_societies}.}

The potential for depolarization, reduced negative campaigning, reduced
partisanship, and reduced ethnic tensions has been a popular argument
for using Instant Runoff Voting (IRV, also known as single-winner
ranked-choice voting, Hare, or the alternative vote) \citep{FairVote,reilly2004}.
However, while IRV incentivizes candidates to care about the opinions
of a greater share of the electorate than Plurality (first-past-the-post)
does, it still incentivizes them to care about some voters more than
others. Consider a simple example with three candidates. Voter 1 has
the preferences A>B>C and might be persuaded that B is better than
A. Voter 2 has the preferences C>A>B and strongly prefers C to either
A or B; this voter might be persuaded to prefer B to A, but won\textquoteright t
be persuaded to prefer either of them to C. Suppose candidate B is
debating which of these voters to reach out to. 
\begin{itemize}
\item Under Plurality there is no benefit to reaching out to Voter 2, who
isn\textquoteright t going to be swayed from voting for C.
\item Under IRV there is a benefit to reaching out to Voter 2: if C is eliminated
first, having Voter 2 vote C>B>A instead of C>A>B will yield an additional
vote for B in the final round. However, reaching out to Voter 1 is
more valuable; doing so can yield B an additional vote in the first
round as well as in a final round against A.
\item Under a Condorcet method such as Minimax, getting Voter 2 to vote
C>B>A is exactly as valuable as getting Voter 1 to vote B>A>C. Either
change yields an additional vote in the pairwise contest between B
and A and is irrelevant in other pairwise contests.
\end{itemize}
This example shows both the advantage of IRV over Plurality and the
potential for other voting methods to provide more equitable incentives
than IRV. Broadly speaking, we expect that when candidates are incentivized
to appeal to a greater fraction of voters, the more likely it is that
candidates will benefit from attracting support outside their most
natural constituency, and so the stronger the centripetal effects
of a voting method will be. The less favorably predisposed a bloc
of voters is towards a candidate, the more likely it is that an incentive
to appeal to those voters will serve a centripetal role. The goal
of this paper is to provide a quantitative description of the electoral
incentives that different voting methods present to candidates seeking
to maximize their chances of winning.

In this paper, we define a voting method\textquoteright s Candidate
Incentive Distribution (CID) as a metric for how valuable it is for
a candidate to reach out to one voter rather than to another, as determined
by the voters\textquoteright{} preexisting attitudes toward that candidate.
As described in Section \ref{sec:Definitions}, a voting method\textquoteright s
CID is determined by generating many simulated electorates, sorting
the voters based on their opinions towards a particular candidate,
perturbing the utilities of some of the voters for the candidate in
question, and seeing if the perturbation results in a different winner.
Cases in which this causes the candidate to go from losing to winning
or from winning to losing are identified as cases where the candidate
has an incentive to appeal more to the voters whose utilities were
perturbed.

While a voting method's CID captures candidates' incentives to appeal
directly to voters, it should be remembered that there can be other
electoral incentives at play. Some of them, such as incentives for
fundraising, recruiting campaign volunteers, and attracting media
attention, seem like they should bear relatively little dependence
on the voting method being used. (This is not certain. For example,
if campaigning spending is more effective at attracting useful support
under some voting methods than others, we should expect a commensurate
difference in incentives related to fundraising.) Had we tried to
account for these additional incentives, we would most likely see
a stronger incentive to appeal to the most supportive voters for all
voting methods since these voters are the best predisposed towards
volunteering or donating. Other incentives can be captured partially
by CID. Incentives to earn support from political parties, such as
on ``how-to-vote'' cards that show a party's suggested preference
order and are commonly used in Australia, are reflected by CID insofar
as parties endorse candidates based on who they think is closest to
them ideologically. However, CID does not capture the possibility
of quid pro quo deals between parties, such as for two parties to
endorse ranking one another's preferred candidate second.

An alternative to centripetalism is the theory of consociationalism,
which says that an ethnically divided society's internal tensions
are best managed with an electoral system that ensures that all ethnic
groups receive representation and that no one group can wield power
without the support of one or more other groups \citep{lijphart1969consociational,lijphart2004constitutional}.
The consociational approach relies on post-electoral negotiations
among elites to manage ethnic tensions, and consociationalists prefer
proportional representation to ensure that all major ethnic groups
are represented. Centripetalists, by contrast, focus on electoral
incentives prior to elections and often prefer single-winner districts
to proportional voting methods under the theory that single-winner
districts can yield stronger centripetal incentives (a claim that
our findings support) \citep{horowitz2001ireland}.

While CID is directly relevant to centripetalism, it says nothing
about how effective a voting method is from a consociationalist perspective.
Therefore, CID cannot answer the question of whether the centripetal
or consociational approach is more effective except insofar as it
can help us interpret empirical research. What we can do is evaluate
the CIDs of proportional voting methods that are effective from a
consociational perspective to see if they provide significant centripetal
incentives as well; we do so in Section \ref{sec:Results-for-multi-winner}.

CID does not tell us whether an incentive to appeal to a particular
group of voters constitutes a centripetal or centrifugal incentive.
If a candidate is not close to the center of public opinion, the most
supportive voters are among the most likely to provide centrifugal
incentives, and the most opposed voters will be opposing extremists,
for whom an incentive to appeal toward will be centripetal. Voters
who like a candidate the median amount are likely to be near the center
of public opinion (though they might also be extremists within the
candidate's faction), and whether or not such voters provide a centripetal
incentive depends on the society in question, as we discuss in the
conclusion.

\section{Related Work\label{sec:Related-Work}}

Much of the existing research on how voting methods differently incentivize
candidates, and on how these incentives influence societies, is in
the context of centripetalism, Reilly \citeyearpar{reilly_divided_societies,reilly2004}
claims that centripetalist systems with IRV or Single Transferable
Vote (STV) have been effective in several divided societies including
Papua New Guinea, Fiji, and Northern Ireland. Of these, Papua New
Guinea offers the closest thing to an unambiguous success story for
these voting methods with ``much stronger local polarization'' \citep{PNG_rumsey1999}
after switching from IRV to Plurality in 1975 and a reduction in electoral
violence when a version of IRV that allowed only three rankings was
introduced in 2002 \citep{PNG_wood,PNG_reilly2021}. Despite being
skeptical of centripetalism more generally, \citet{doesmoderationpay}
writes that ``the use of centripetal electoral rules has made {[}Papua
New Guinea{]} more stable than it was previously.'' The effects in
other societies are more controversial. In Fiji, the outcome of the
first IRV election led to a military coup, and the second IRV election
saw moderate parties win far fewer seats than would be expected based
on their shares of first preferences, with more votes being transferred
from moderate to extremist parties than vice versa \citep{arms2006Fiji,coakley2017ethnic}.
Regarding Northern Ireland, a major question is whether centripetal
effects of STV played a role in bringing about a lasting peace. \citet{irelandstv}
notes that ``Inter-ethnic vote-pooling was virtually non-existent''
before the 1998 Belfast agreement, but became non-negligible afterward.
While \citet{reilly2004} describes the incentive to secure lower-order
preferences as ``instrumental'' to Sinn Fein and other parties becoming
less extreme, \citet{bogaards2019friendsorfoes} describes the post-agreement
system as ``full-fledged consociational'' and writes that we ``cannot
tell where political moderation comes from, whether it is due to a
centripetal electoral system and/or a consociational grand coalition.''

Prior work employing computer simulations to study electoral incentives
has used candidate positioning models to assess the centripetal or
centrifugal effects of voting methods. \citet{plurality_centrifugal_incentives}
use a partisan voter model to find that, under Plurality, candidates
are often incentivized to take more extreme positions to appeal more
to their core base of support, but do not study the incentives under
other voting methods. \citet{atsusaka_landsman} use a two-dimensional
model with three factions, where each faction has both a moderate
and an extremist party, to analyze the differing incentives under
IRV. They consider three possibilities: Plurality (FPTP), IRV where
parties/candidates try to win both first- and second-choice support,
and IRV where parties and candidates try to win first-, second-, and
third-choice support. They find that ``switching from FPTP to RCV
{[}IRV{]} makes a large difference only when parties are incentivized
to maximize not only their second-choice probabilities but also their
third-choice probabilities''. However, they do not evaluate how valuable
this second- and third-place support is to candidates, so their model,
by itself, does not yield a clear answer as to whether IRV and Plurality
yield substantially different incentives. \citet{robinette2023implications}
uses a one-dimensional voter model and trains a neural network to
predict how likely candidates are to win at different ideological
positions, based on the positions of other candidates. He finds that
ideologically strategic candidates will avoid the exact center of
public opinion under IRV, but will converge to it under Condorcet
methods.

While Merrill, Adams, Atsusaka, and Landsman all model candidates/parties
as maximizing their vote shares, leaving open the question of what
exactly constitutes a vote, CID, like \citet{robinette2023implications},
models candidates as maximizing their win probabilities as suggested
by \citet{stoffel2014unified}. The extent to which candidates are
incentivized to compete for second-choice support vs. third-choice
support is baked into the model: candidates will value each kind of
support in proportion to how much it helps them win. This also means
that CID can be used to evaluate any voting method, rather than being
limited to Plurality and IRV, and that it accounts for differences
in incentives that arise from different voting methods using different
tabulation algorithms with identical ballots.

Our approach is different from and complementary to that of candidate
positioning models. CID and candidate positioning models provide answers
to slightly different questions. CID does not directly tell us where
winning candidates will lie on an ideological spectrum; instead, it
tells us which voters pull candidates the most strongly toward their
positions. What CID tells us directly, that candidate positioning
models do not, is how much more important some voters are than others
from the perspective of a candidate maximizing their odds of being
elected. Unlike candidate positioning models, CID is not tied to any
particular spatial voter model and can be used with non-spatial models
if desired.

Other related research includes differences in campaign civility under
Plurality and IRV. Sentiment analysis of both mayoral debates \citep{civildebates}
and campaign communications \citep{campaigncommunications} shows
more positive campaigning under IRV than under Plurality, and polls
of voter perceptions further support this claim \citep{civilitypolls}.
Such findings of more civil campaigning are relevant to centripetalism.
As \citet{Reilly_established_democracies} notes, \textquotedblleft The
hope that greater political civility and cooperation can be fostered
by making politicians responsible to a broader electorate via changing
the rules of the electoral game mirrors the objective of \textquoteleft making
moderation pay\textquoteright{} for centripetal reformers in ethnically
divided societies.\textquotedblright{}

However, reducing negative campaigning is not synonymous with reducing
inter-group animosity. There are multiple varieties of negative campaigning
which have different consequences for society. Consider the difference
between a Republican in the United States dubiously accusing another
Republican of being corrupt and dubiously accusing a Democrat of the
same thing. The latter is far more likely than the former to exacerbate
societal cleavages, and it is only the incentive for the latter (or
more precisely, the lack of an incentive to avoid further alienating
opposing voters by doing the latter) that is captured to a significant
extent by CID. 

\section{Definitions\label{sec:Definitions}}

Let $n$ be the number of voters, $m$ be the number of candidates,
$w$ be the number of winners, $W$ be the set of all $\binom{m}{w}$
possible sets of winners, and $P$ be the space of all possible polls\footnote{For all of these definitions, a poll can be an arbitrary mathematical
object. However, the only polls we consider are elements of $\left[0,1\right]^{m}$,
i.e. polls that give a single number describing the support for each
candidate.}.
\begin{defn}
A \emph{voting method} consists of a set of valid ballots $B_{f}$
and a tabulation function $f:B_{f}^{n}\rightarrow W$.
\end{defn}

\begin{defn}
A \emph{voter} $v$ is a vector in $\mathbb{R}^{m}$, where $v_{j}$
is the utility\footnote{A \emph{utility} is a real number that describes how good a voter
thinks a candidate is. Candidates who are assigned higher utilities
are considered better.} assigned to the $j$th candidate. The set of all possible voters
is denoted $V$. An \emph{electorate} $G$ of $n$ voters is an $n\times m$
matrix in which every row is a voter. A \emph{voter model} is a probability
distribution over electorates.
\end{defn}

\begin{defn}
A \emph{strategy} is a function $t:(V\times P)\rightarrow B_{f}$.
Given an electorate $G$ of voters $v_{1},v_{2},...,v_{n}$ and a
vector of strategies $\vec{t}=\left(t_{1},t_{2},...,t_{n}\right)$
we write

\[
\vec{t}(G,p)\coloneqq\left[\begin{array}{c}
t_{1}(v_{1},p)\\
t_{2}(v_{2},p)\\
...\\
t_{n}(v_{n},p)
\end{array}\right]
\]
\end{defn}

To determine whether a candidate is incentivized to appeal to a particular
group of voters, we slightly increase or decrease those voters' opinions
of the candidate and see if this turns the candidate from a loser
to a winner or from a winner to a loser. We must also aggregate the
results of this across different candidates and different electorates.
To set up the definition of a Candidate Incentive Distribution (CID),
we need to define a few more symbols:
\begin{itemize}
\item $c$ is the index of a candidate, so $1\leq c\leq m$. We will have
$c$ be a random variable that is uniformly distributed over all candidates
in order to aggregate the incentives across different candidates.\footnote{An alternative approach is to let CID be a function of $c$ and to
use an asymmetric voter model that treats some candidate indices differently
from others. We do not consider asymmetric voter models in this paper.}
\item $S_{c;G}$ is an $n\times n$ permutation matrix that, when applied
to $G$, sorts the electorate based on their opinions of candidate
$c$, with its inverse given by $S_{c;G}^{T}$. These matrices can
also be applied to vectors of ballots or to electorates other than
$G$. We denote a sorting function that creates $S_{c;G}$ from $c$
and $G$ as $S$. Such a sorting function is discussed later.
\item $p(G,\xi)$ is a polling function whose inputs are an electorate and
a random variable $\xi$, which represents the polling noise. It maps
to $P$.
\item $H$ is a voter model, and $G$ is a random matrix that is distributed
according to $H$.
\end{itemize}
After sorting voters according to some $S$, we group voters into
equal-sized buckets by letting $K$ be the number of buckets (which
divides $n$) and $k\in\left\{ 1,2,...,K\right\} $ be an index that
specifies a particular bucket. Then, given some number $\epsilon>0$,
let $\epsilon^{k,c}$ be a matrix that, when added to $S_{c;G}G$,
increases the utilities that voters in the $k$th bucket assign to
candidate $c$ by $\epsilon$. Explicitly, $\epsilon^{k,c}$ is an
$n\times m$ matrix with
\[
\epsilon_{i,j}^{k,c}\coloneqq\begin{cases}
\epsilon & \text{if }\left(k-1\right)\cdot\frac{n}{K}<i\leq k\cdot\frac{n}{K}\text{ and }j=c\\
0 & \text{otherwise}
\end{cases}
\]

\begin{defn}
\label{def:UCI}The \emph{Unnormalized Candidate Incentive} is

\begin{align*}
 & UCI\left(f,\vec{t},n,m,H,\epsilon,S,p,K,k\right)=\\
 & \quad\Pr\left[c\in f\left(\vec{t}\left(G,p(G,\xi)\right)\right),c\notin f\left(\vec{t}\left(S_{c;G}^{T}\left(S_{c;G}G-\epsilon^{k,c}\right),p(G,\xi)\right)\right)\right]\\
 & +\Pr\left[c\notin f\left(\vec{t}\left(G,p(G,\xi)\right)\right),c\in f\left(\vec{t}\left(S_{c;G}^{T}\left(S_{c;G}G+\epsilon^{k,c}\right),p(G,\xi)\right)\right)\right]\\
 & -\Pr\left[c\in f\left(\vec{t}\left(G,p(G,\xi)\right)\right),c\notin f\left(\vec{t}\left(S_{c;G}^{T}\left(S_{c;G}G+\epsilon^{k,c}\right),p(G,\xi)\right)\right)\right]\\
 & -\Pr\left[c\notin f\left(\vec{t}\left(G,p(G,\xi)\right)\right),c\in f\left(\vec{t}\left(S_{c;G}^{T}\left(S_{c;G}G-\epsilon^{k,c}\right),p(G,\xi)\right)\right)\right]
\end{align*}
\end{defn}

In the first term, $f\left(\vec{t}\left(G,p(G,\xi)\right)\right)$
is the set of winners when voters' utilities are not perturbed. $S_{c;G}G-\epsilon^{k,c}$
is the electorate matrix with the voters sorted according to $S$
and the utilities that the voters in the $k$th bucket assign to $c$
reduced by $\epsilon$. $\vec{t}\left(S_{c;G}^{T}\left(S_{c;G}G-\epsilon^{k,c}\right),p(G,\xi)\right)$
is the vector of ballots cast with the perturbed utilities; the purpose
of $S_{c;G}^{T}$ is to ``unsort'' the voters such that they use
the same strategies regardless of whether their utilities are perturbed
or not. Applying $f$ to this yields the set of winners with reduced
utilities for $c$. Thus, the first term is the probability that a
randomly chosen candidate will win in the baseline scenario but will
lose if the voters in the $k$th bucket like them modestly less. Similarly,
the second term is the probability that a randomly chosen candidate
will lose in the baseline scenario but will win if the voters in the
$k$th bucket like them modestly more.

The third term represents the unlikely scenario in which a candidate
being more liked by some voters causes them to lose. This means that
the candidate is better off if these voters are alienated, all else
being equal, so this term is negative. The fourth term is analogous.
The absolute value of these terms is much lower than that of the first
two terms, and for many voting methods is exactly zero.

All of these probabilities are approximated by running thousands of
computer simulations. In each iteration, an electorate is generated
(i.e. randomly sampled) from the voter model $H$ and the polling
is conducted. Then, for each voting method and strategy under consideration,
the outcome is determined in the baseline case with unperturbed utilities.
For each candidate, the voters are sorted according to $S_{c;G}$,
and, for each bucket, the utilities of the voters in that bucket are
perturbed. These voters cast new ballots, and the winners are determined
again.

We also want to normalize the Unnormalized Candidate Incentive. This
is partly to minimize the importance of $K$ and $\epsilon$. Normalizing
also means that differences in the frequency of close elections under
different voting methods do not affect CID, so as to better compare
the relative incentives to appeal to some voters rather than others.
\begin{defn}
The \emph{Candidate Incentive} is 
\[
CI\left(f,\vec{t},n,m,H,\epsilon,S,p,K,k\right)=\frac{K\cdot UCI\left(f,\vec{t},n,m,H,\epsilon,p,S,K,k\right)}{\sum_{k'=1}^{K}UCI\left(f,\vec{t},n,m,H,\epsilon,p,S,K,k'\right)}
\]

This normalization means that the average Candidate Incentive, across
all buckets, is always one. A Candidate Incentive of 2 for the $k$th
bucket means that candidates are (on average) twice as strongly incentivized
to appeal to a voter in the $k$th bucket as to the average voter,
and a Candidate Incentive of $1/10$ means that candidates should
be indifferent between increasing their appeal by $\epsilon$ to ten
voters in the $k$th bucket and increasing their appeal by $\epsilon$
to a single voter randomly chosen from among all buckets. Fixing everything
except $k$ and letting $x=\left(k-1\right)/\left(K-1\right)\in\left[0,1\right]$,
the \emph{Candidate Incentive Distribution} (CID) is the graph of
the candidate incentive with respect to $x$. We will use $x$ for
the charts (expressed as a percentage) and switch between using $k$
and $x$ depending on convenience.
\end{defn}

Note that the Unnormalized Candidate Incentive is defined such that
the perturbations of voter utilities do not affect the polls. This
is partly to reduce the dependence of CID on a somewhat arbitrary
choice of polling function and partly to reduce the time required
to run simulations. However, this choice means that CID does not capture
the value of appealing to a voting bloc for the sake of getting better
numbers in the polls; this is especially important for Plurality elections,
where doing well in the polls reliably increases support from strategic
voters.

In this paper we focus on the dependence of CID on $f$ (the voting
method), $\vec{t}$ (the strategies), and $m$ (the number of candidates).
The role of other parameters is explored in Appendix \ref{sec:Appendix-2:-Additional-results}.
For the main paper:
\begin{itemize}
\item $H$ is the clustered spatial model used in \citet{wolk2023star}.
This model yields a wide variety of electorates with different numbers
of clusters and issue dimensions, with the number of dimensions and
the extent to which voters care about each dimension being determined
by a Dirichlet stick-breaking process. A similar Chinese Restaurant
process is also used to assign voters to clusters. Simpler spatial
models yield similar results, as shown in Appendix \ref{sec:Appendix-2:-Additional-results}.
\item $n$ is 72 and $K$ is 24. These values were chosen to minimize distortions
from a small electorate and to show differences in incentives at a
relatively high resolution without requiring excessive computational
time.
\item $\epsilon$ is approximately 11\% of a standard deviation of a voter's
candidate utilities (averaged across all voters generated by $H$).
For five-candidate elections with ranked ballots, this corresponds
to being ranked an average of 0.17 places higher, a score that is
higher by an average of 0.22 under STAR, and receiving an additional
vote 5.1\% of the time under Approval Voting. These numbers were determined
experimentally after the value of $\epsilon$ was set.
\item The sorting function $S$ works as follows: First, each voter's utilities
are rescaled to have mean 0 and standard deviation 1. Then the voters
are sorted by their rescaled utilities for the $c$, in order of increasing
utility.
\item The polling function used depends on the voting method under consideration.
Polling is discussed in section \ref{sec:Results-with-strategic}.
\end{itemize}
In addition to CID, we define a summary metric to capture how uneven
or lopsided a CID is. A voting method's CID can be interpreted as
a probability distribution\footnote{Specifically, given that an election would have had a different outcome
for a candidate if a voter had a slightly different opinion of that
candidate, the CID (in the limit $n\rightarrow\infty$ and $K\rightarrow\infty$)
is the probability density of the voter having some preexisting attitude
toward that candidate. This assumes the last two terms in Definition
\ref{def:UCI} are negligible.}, which suggests that a metric from probability theory may be a good
descriptor. To capture the intuition that a CID which is 0 for $x<1/2$
and 2 for $x>1/2$ is more lopsided than a CID which is 2 for $1/4<x<3/4$
and zero otherwise, we employ the earth mover's distance to show how
much and how far probability mass must be shifted to get the uniform
distribution.
\begin{defn}
Given a CID $C\left(k\right)$ which uses $K$ buckets, the \emph{earth
mover's distance from uniformity} (EMU) is
\[
EMU\left(C\right)=\frac{1}{2K^{2}}\min_{\mu}\left(\sum_{k,l}\left|\mu_{kl}\right|\left|k-l\right|\middle|\forall k:C(k)+\sum_{l=1}^{K}\mu_{kl}=1,\mu_{kl}=-\mu_{lk}\right)
\]

Here, $\mu_{kl}$ denotes the probability mass\footnote{Normalized to $K$ rather than 1}
to be shifted from bucket $l$ to bucket $k$.
\end{defn}

\section{Results with sincere voters\label{sec:Results-with-non-strategic}}

We consider the following single-winner voting methods:
\begin{itemize}
\item Plurality: Voters vote for only one candidate. Whoever receives the
most votes wins.
\item Approval: Voters vote for any number of candidates. Whoever receives
the most votes wins.
\item Plurality Top 2: In the first round, voters vote for only one candidate.
The two candidates with the most votes advance to a runoff in which
each voter votes for a single candidate, and the finalist with the
most votes in the runoff wins.\footnote{While real-world uses of Plurality Top 2 involve two separate elections,
here we implement it as a single election in which a ballot is a specification
of who receives a vote in the first round together with a separate
complete ranking of candidates; the higher-ranked finalist receives
a vote in the runoff. This means that voter preferences remain constant
between the first round and the runoff, but voters may vote strategically
such that their top-ranked candidate does not receive their vote in
the first round. Our model does not account for the possibility of
decreased turnout in the runoff among voters who dislike both finalists,
as has been found by \citet{orphanedvoters}.}
\item Approval Top 2: Same as Plurality Top 2, except that voters can vote
for any number of candidates in the first round.
\item Instant Runoff Voting (IRV): Each ballot is a ranking of some or all
of the candidates.\footnote{We do not model scenarios in which there are fewer allowed rankings
than there are candidates, and none of the strategies we consider,
aside from bullet voting, involve ballot truncation.} At the beginning of tabulation, each ballot is counted as one vote
for its highest-ranked candidate. In each round, the candidate with
the fewest votes is eliminated and their votes are transferred to
the highest remaining choices on their supporters' ballots. The last
candidate remaining wins.
\item STAR Voting: Voters score the candidates from 0 to 5. The two candidates
with the highest total scores are finalists, and whichever finalist
receives a higher score than the other on more ballots wins.
\item Minimax: A Condorcet method.\footnote{Condorcet cycles occur in less than 2\% of elections generated from
our voter model, so all Condorcet methods have a very similar CID
to Minimax.} As with IRV, each ballot is a ranking of some or all of the candidates.\footnote{While it is often recommended that equal rankings be allowed under
Minimax \citep{darlington2022condorcet}, we do not consider such
strategies in this paper.} In the event of a Condorcet cycle, the winner is the candidate whose
largest margin of head-to-head defeat is the smallest.
\end{itemize}
Ties are resolved in favor of whichever candidate has the lowest index.
We chose not to evaluate Borda Count, Coombs' Method, or Anti-Plurality
because of their erratic behavior in the presence of strategic voting.

As our baseline, we consider elections with 5 candidates\footnote{This is partly due to the popularity of Final Five Voting, a reform
primarily aimed at improving candidate incentives \citep{finalfive}.} and all voters behaving sincerely without the use of any polling
data. For all ordinal voting methods there is a unique sincere strategy
(so long as voters do not assign multiple candidates equal utilities),
but sincere voting needs to be defined for Approval, Approval Top
2, and STAR. For Approval and Approval Top Two, the sincere\footnote{The use of the term ``sincere'' should not be taken to imply that
the other strategies we consider for cardinal voting methods are insincere.
For cardinal voting methods, every strategy we consider in this paper
is sincere in the sense that a voter will virtually never give greater
support to Candidate A than to Candidate B unless she prefers A to
B.} strategy is to rescale one's utilities to a scale such that one's
favorite candidate is assigned a utility of 1 and the average utility
of the candidates is 0, and to vote for all candidates whose utility
is at least 0.4\footnote{This value was tuned to be approximately strategically optimal within
the constraints of the parameterized strategy, as shown in Appendix
\ref{sec:Appendix-1:ESIF}.}.\footnote{It is implausible that many voters would make the explicit calculations
for using this strategy; realistically, voters will use difficult-to-articulate
heuristics for determining their approval thresholds. This strategy
should be thought of as an algorithm that approximates realistic voter
behavior rather than as a description of realistic voter thought processes.
Tuning the parameter for strategic optimality yields a reasonable
best guess as to voter behavior and reduces the opportunity for experimenter
bias.} For STAR, the sincere strategy is to rescale one's utilities to a
0-5 scale\footnote{The linear rescaling is such that one's favorite candidate is assigned
5 utility. The unscaled utility that is scaled to 0 is the minimum
of the utility of one's least favorite candidate and the utility of
the average candidate, minus the difference in utility between one's
favorite candidate and the average candidate.} and give each candidate a score equal to the utility assigned to
them, rounded to the nearest whole number.

\begin{figure}
\includegraphics[scale=0.7]{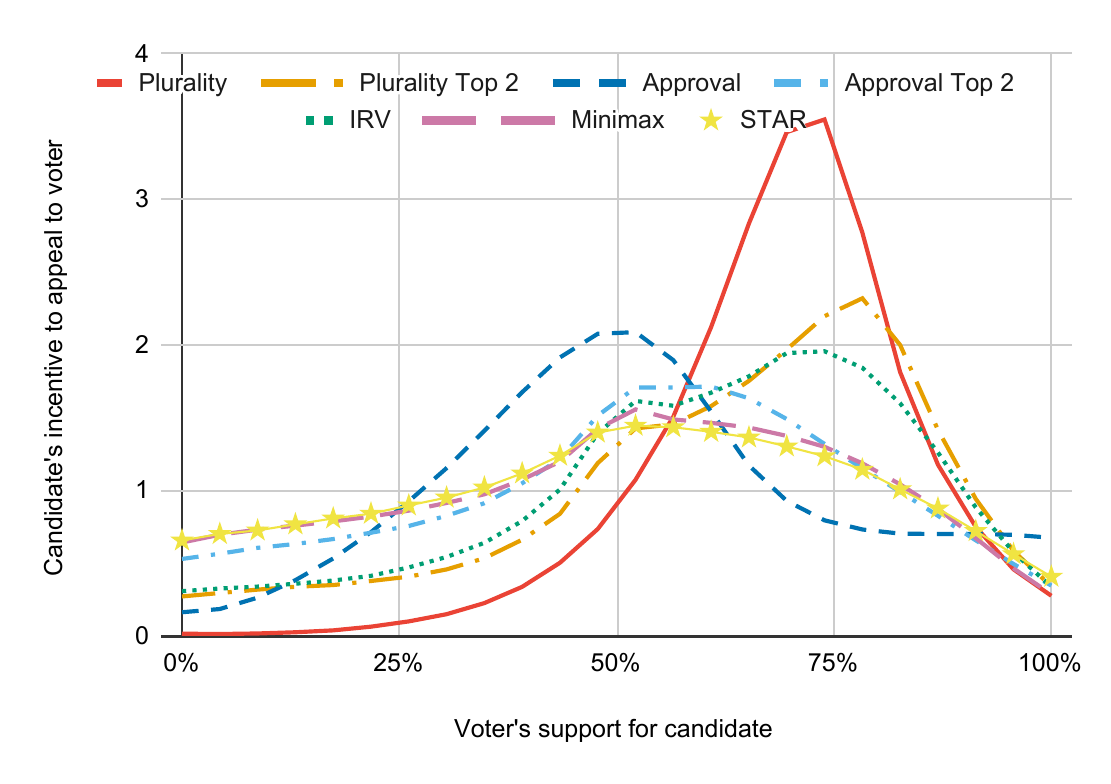}\caption{\label{fig:hon5c}Sincere voters, 5 candidates, 250,000 iterations}
\end{figure}

We can see in figure \ref{fig:hon5c} that the CIDs of these voting
methods differ most notably in three respects: The height of the peak,
the location of the peak, and their behavior as $x\rightarrow0$.

Plurality has by far the highest peak, and this peak is well to the
right of 50\% since candidates can usually win with significantly
less than 50\% of the vote. Towards the left of the graph, Plurality's
CID drops almost to zero; if a voter can't be persuaded to prefer
a candidate above all others, the candidate has no incentive to avoid
alienating them.

Plurality Top 2 does offer a meaningful incentive not to alienate
already-opposed voters with a Candidate Incentive of around 1/3 toward
the left side of the chart. It has a lower peak than Plurality, but
this peak is slightly to the right of Plurality's. This is because
the goal in Plurality is to get the most votes, but the goal in the
first round of Plurality Top 2 is to be one of the two candidates
with the highest vote totals, which implies a slightly lower threshold
of needed support than Plurality. However, this is a rather minor
effect; the addition of the runoff results in a far less lopsided
CID.

IRV's CID is very similar to that of Plurality Top 2, but is slightly
higher in the 30-60\% region and has a correspondingly lower peak.
However, the difference becomes negligible towards the left of the
chart; the votes there are unlikely to be transferred to the candidate
prior to going to the ``lesser of two evils'' in the final round.

Minimax's CID is very different from those of the other ordinal methods;
its (much flatter) peak is in the center, and it provides a far greater
incentive to appeal to opposing voters; Minimax's Candidate Incentive
for the most strongly opposed voters is .64, compared to .31 for IRV.

It is worth considering a basic question about Minimax. If a voter
slightly prefers Candidate A to Candidate B, the advantage to Candidate
B of convincing this voter to prefer her to Candidate A is the same
whether candidates A and B are preferred to all other candidates,
liked less than all other candidates, or together in the middle of
the pack in this voter's opinion. Given this, why is Minimax's CID
not perfectly uniform? The answer is that Minimax's CID tells us the
approximate distribution of voters who are indifferent between the
Condorcet winner and the Condorcet runner-up --- and, under our voter
model, such voters are the most likely to have an opinion of these
candidates that is slightly more positive than that of the median
voter.

Of the cardinal methods, Approval has the highest peak. This is a
simple consequence of Approval, like Plurality, only offering two
levels of support. For both of these voting methods, the peak lies
at the approximate threshold between supporting or not supporting
the candidate. However, Approval's peak lies at the center rather
than towards an extreme. Approval does incentivize candidates to care
about the preferences of some voters substantially more than others,
but under Approval Voting these especially influential voters tend
to have a middling view of that candidate, whereas with Plurality
they view the candidate quite favorably. Approval also offers a meaningful
(albeit lesser) incentive to appeal to voters with an extreme opinion
of the candidate. This is because even though these voters are unlikely
to change their minds on whether to vote for the candidate in question,
liking this candidate more will make them less inclined to vote for
other candidates.

As with Plurality and Plurality Top 2, adding a runoff to Approval
Voting increases the incentives to appeal to relatively opposed voters
in order to win their vote in the runoff against an also-disliked
challenger. However, this incentive appears to be stronger in Approval
Top 2 since the finalists under Approval Top 2 tend to be similar
to one another (since they have to be at least somewhat close to the
political center to make it that far), while under Plurality Top 2
they can easily be opposing extremists. Under Approval Top 2, winning
the vote of a strongly opposed voter in the runoff is a more realistic
prospect\footnote{The claim that runoffs do more to promote moderation when the finalists
are relatively similar is supported by evidence from the Top Two primary
system of California and Washington, which only appears to yield more
moderate winners when the finalists are from the same party \citep{top2sameparty}.}.

STAR is similar to Approval Top 2 in that both are cardinal voting
methods that include a runoff, but STAR's CID is more even than Approval
Top 2's since STAR offers more than two levels of support. STAR's
CID is very similar to Minimax's, but slightly more even. The most
natural explanation for the additional evenness is that our sincere
strategy for STAR allows voters to give the same score to both finalists,
while our strategy for Minimax does not involve tied rankings. Therefore,
for the pairwise comparison between these candidates (as occurs in
both STAR's runoff and the tabulation of Minimax), STAR gives voters
three options but Minimax only gives two, and the additional relevant
threshold of support means that STAR's CID has less of a peak. 

Even within the constraints of not using polling data and voting sincerely,
voter strategy plays a major role in the CIDs of cardinal voting methods
since the overall willingness of voters to give significant support
to multiple candidates determines the location of a CID's peak. If,
under any given cardinal voting method, every voter were to bullet
vote (give their favorite candidate full support and no support to
anyone else), its CID would be the same as Plurality's (or the CID
of Plurality Top 2, in the case of Approval Top 2). Alternatively,
if every voter were to do the opposite of bullet voting (giving full
support to every candidate except their least favorite), the peak
would be far on the left side of the chart.

However, strategically reasonable behavior, even in the absence of
information about the preferences of other voters, requires that the
CIDs of these voting methods must peak approximately in the center.
Voters want to have influence, and if the peak of the CID one of these
voting methods was at $x$ significantly greater than 50\%, it would
mean that voters who were more likely to vote for any given candidate
would, on average, wield greater influence. This means that voters
would be incentivized to support more candidates, pulling the peak
of the CID to the left. Likewise, a peak at $x$ much less than 50\%
would suggest an opportunity to gain influence by supporting fewer
candidates. Only a peak near 50\% does not allow for these easy strategic
optimizations. 

That said, the sincere STAR strategy has a central peak without any
numerical fine tuning\footnote{Its peak would be somewhat towards the left under a simpler strategy
in which a voter's last choice always received a rescaled utility
of 0 and the utility of the mean candidate was not considered.}. The fact the peak of Approval Top 2's CID is to the right of Approval's
suggests that voters are incentivized to vote for slightly more candidates
if there is a runoff\footnote{This is confirmed in Appendix \ref{sec:Appendix-1:ESIF}. The decision
to use the same strategy for Approval and Approval Top 2 was made
for the sake of an apples-to-apples comparison.}.

\begin{figure}
\includegraphics[scale=0.7]{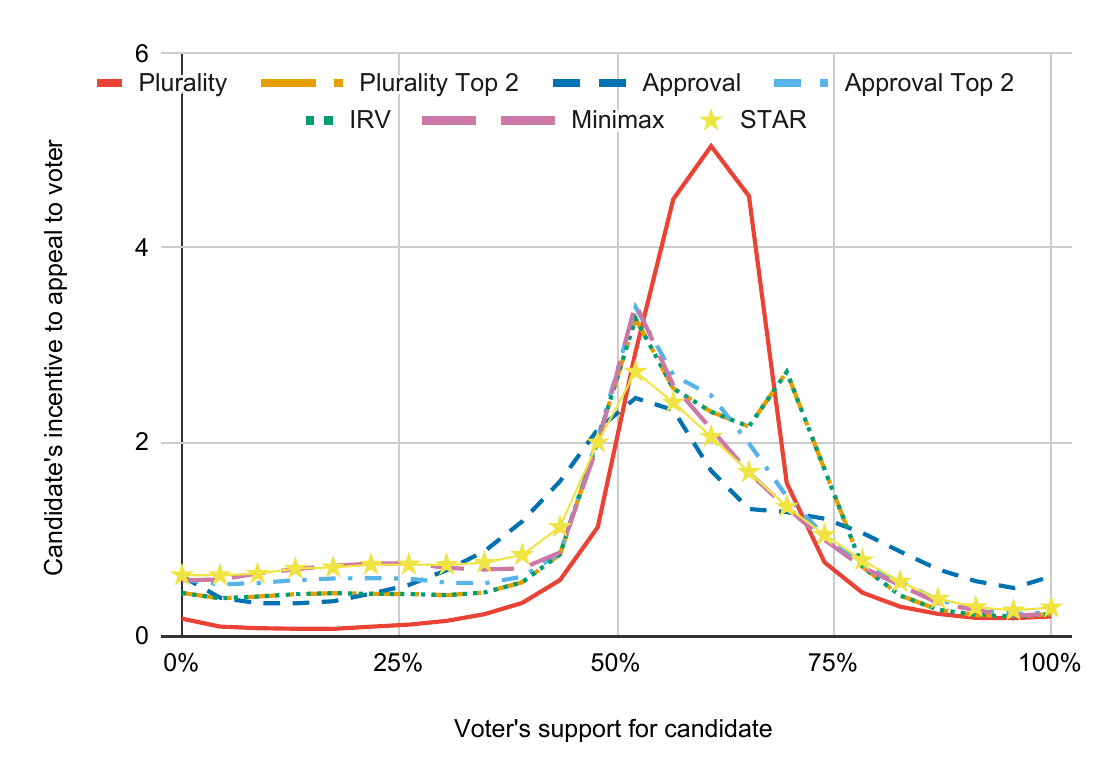}\caption{\label{fig:hon3c}Sincere voters, 3 candidates, 250,000 iterations}

\end{figure}

\begin{figure}
\includegraphics[scale=0.7]{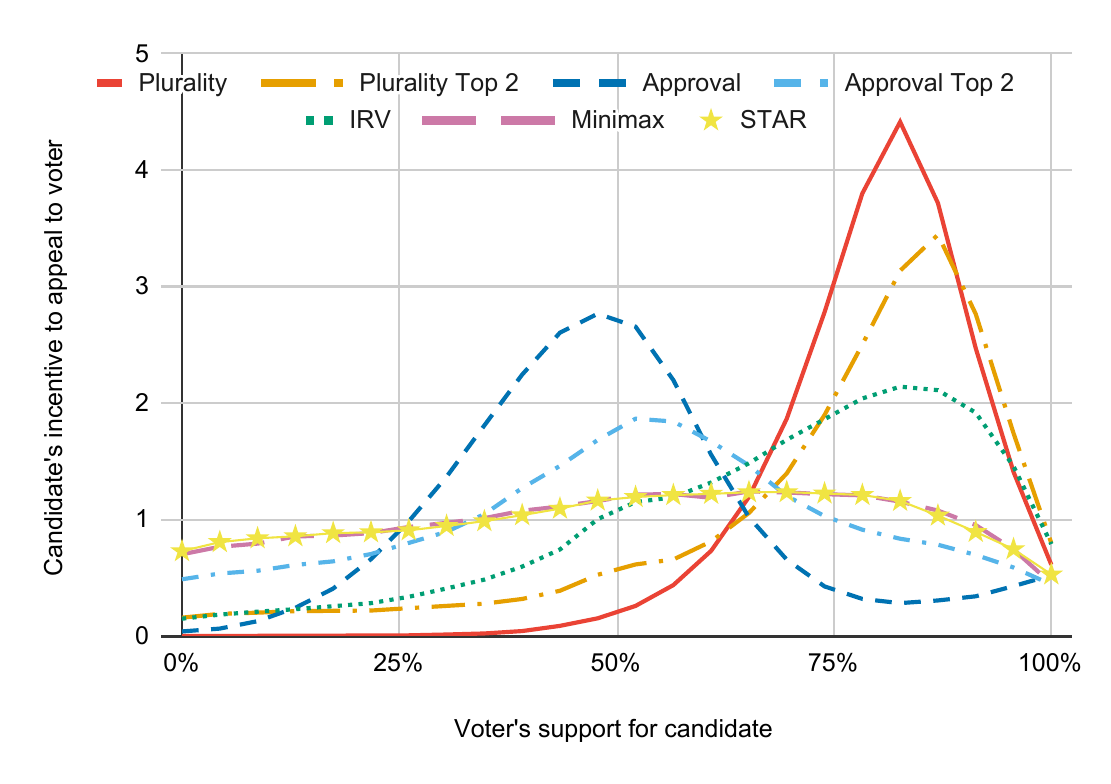}\caption{\label{fig:hon10c}Sincere voters, 10 candidates, 100,000 iterations}

\end{figure}

We can also consider elections with more or fewer candidates (Figures
\ref{fig:hon3c} and \ref{fig:hon10c}). With three candidates, every
voting method except Plurality has a peak in the center. Plurality
Top 2 and IRV behave identically and have two peaks, the one on the
right corresponding to the incentive to have enough first-place support
to advance to the runoff and the central peak corresponding to the
incentive to have enough first- and second-place support to win the
runoff. This central peak is also apparent in the five-candidate chart,
but there it is far less pronounced and is more of a shoulder. As
the number of candidates in a race increases, so does the importance
of having first-place supporters in Plurality Top 2 and IRV.

Minimax, STAR, and Approval Top 2 have sharper peaks with three candidates
because the sorting function, which considers how the utility of the
candidate in question compares to the utility of the mean candidate,
is affected more by the strongest rival of the candidate in question
when there are fewer candidates to contribute to the mean. Under all
of these voting methods, candidates have the strongest incentives
to appeal to voters who are approximately indifferent between them
and their strongest rival, and the contribution of this rival to the
mean causes these voters to be clustered near $x=50\%$. Interestingly,
Approval has a more even CID than Approval Top 2 in three-candidate
elections.

With ten candidates, the CIDs of Approval Top 2, STAR, and Minimax
are considerably more even. By contrast, the CIDs of Plurality, Plurality
Top 2, and IRV\footnote{This chart may overstate the degree to which IRV incentivizes candidates
to appeal to opposed voters in many-candidate elections. Real-world
implementations of IRV often limit the number of candidates that voters
are allowed to rank, but in our model we assume that voters rank every
candidate. Minimax faces a similar issue since it also uses ranked
ballots, but if voters are allowed to rank candidates equally the
impact on candidate incentives should be minimal.} are more lopsided; their peaks move further to the right as the number
of candidates increases, and the incentive to appeal to the most opposed
voters decreases commensurately. The difference is particularly acute
for Plurality Top 2; with five candidates its CID is very similar
to IRV's, but with ten candidates its peak is almost as pronounced
as Plurality's. The more candidates there are, the greater the significance
of IRV's intermediate rounds.

\begin{figure}
\includegraphics[scale=0.7]{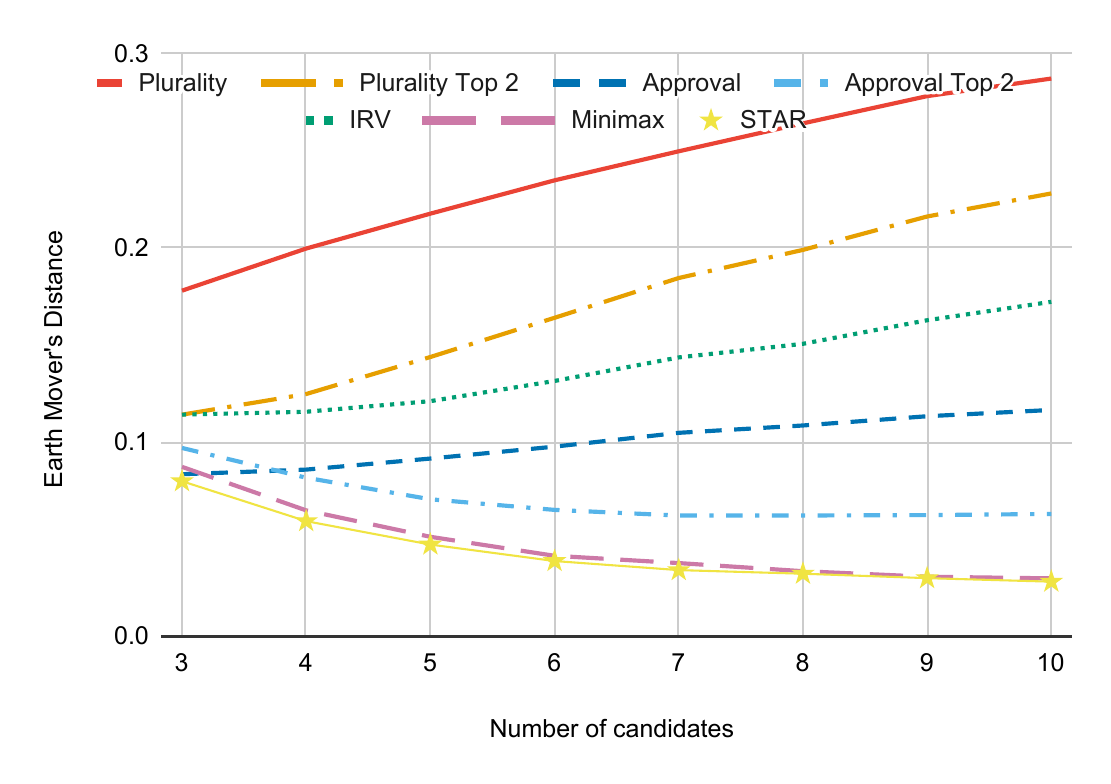}\caption{\label{fig:Earth-Mover's-Distance}Earth Mover's Distance from Uniformity,
100,000 iterations}

\end{figure}

We can see the effects of having more or fewer candidates more clearly
in figure \ref{fig:Earth-Mover's-Distance}. In Plurality, Plurality
Top 2, IRV, and Approval, the incentives become less equitable with
more candidates. This is consistent with Merrill and Adams' \citeyearpar{plurality_centrifugal_incentives}
finding that, under Plurality, ``the more candidates there are, the
more extreme the candidate's optimal position''. With Approval Top
2, STAR, and Minimax, having more candidates makes the incentives
more equitable. Regardless of the number of candidates, Plurality's
incentives are the most imbalanced, followed by Plurality Top 2, IRV,
and then everything else. However, the relative magnitudes of these
differences change considerably with the number of candidates. With
three candidates, Plurality has an earth mover's distance from uniformity
(EMU) of .178, IRV has an EMU of .114, and STAR has an EMU of .080;
the difference between IRV and Plurality is bigger than the difference
between IRV and STAR. With 10 candidates these numbers are .287, .172,
and .028, respectively; IRV is closer to Plurality than to STAR.

\section{Results with strategic voters\label{sec:Results-with-strategic}}

A simulation-based approach cannot, by itself, tell us how voters
will behave in the real world. To account for this uncertainty, we
provide results for a variety of possible voter behaviors. The sincere
strategy model used for Figures \ref{fig:hon5c}-\ref{fig:Earth-Mover's-Distance}
serves as a reasonable baseline, but many voters take strategic considerations
into account when casting their ballots. For example, Eggers and Vivyan
\citeyearpar{whovotesstrategicallyinbritain} found that roughly 35\%
of UK voters were strategic and \citet{Japan_strategic} found that
63 to 85 percent of Japanese voters were strategic. There is additional
research showing substantial amounts of strategic voting under two-round
systems \citep{France_strategic,Brazil_strategic_voting} and primary
elections \citep{primary_strategy}. However, research on the prevalence
(or manner) of strategic voting under IRV, STV, Approval, and STAR
is virtually nonexistent aside from laboratory experiments such as
\citet{dellis2023IRVstratlab}. Here we show the effect of strategic
voting on CID by first considering elections in which all voters behave
strategically and then the EMU with different fractions of voters
behaving strategically. The possibility of voters deviating from sincere
voting non-strategically by exclusively supporting their favorite
candidate, regardless of their more nuanced opinions, is considered
in Appendix \ref{sec:dogmatic-bullet-voting}.

Approaches to modeling strategic voting have included simple and not-so-simple
rules for particular voting methods (such as voting for one's favorite
of the two top-polling candidates under Plurality) \citep{strategy_lab},
and explicit calculations for maximizing utility \citep{eggers2021susceptibility}.
The former approach is a poor choice for providing apples-to-apples
comparisons between voting methods when the choice of strategic rules
is largely arbitrary. The latter approach is theoretically appealing,
but difficult to implement for elections with many candidates and
a wide range of voting methods. It also assumes far greater competence
at strategic voting than is realistic. In two-round systems, for example,
if one's preferred candidate is expected to easily get the most votes
in the first round it can be advantageous to strategically vote for
a ``pushover'' candidate to improve the chances of that favorite
candidate in the runoff. However, studies on voter behavior do not
show significant numbers of voters employing such ``pushover'' strategies
\citep{France_strategic,Brazil_strategic_voting}. We therefore adopt
the intermediate approach of selecting a rule for each voting method
based on the following universal principles, employing viability-aware
strategies as were used in \citet{wolk2023star}:
\begin{itemize}
\item Only polls that give a single number for each candidate, rather than
something like a pairwise comparison matrix or a full IRV tabulation,
may be used.
\item Polling uncertainty must be handled robustly; if Candidate A is leading
Candidate B in the polls by 1\%, this should not be considered vastly
different from Candidate B leading Candidate A by 1\%.
\item The strategies should be as intuitive and conceptually simple as possible.
They should align with how voters actually think about voting strategically
while being expressed as equations or algorithms rather than as vague
heuristics.
\item Within these constraints, the strategies should be as effective as
possible from the perspective of the voters employing them; assessing
the effectiveness of a strategy is discussed in Appendix \ref{sec:Appendix-1:ESIF}.
\end{itemize}
These strategies depend on an estimate of each candidate's probability
of winning\footnote{This is as described in \citet{wolk2023star}, ``the probability
of being truly in first place is calculated by assuming that each
candidate's true percent approval is independently beta-distributed
around their observed approval, tuning the total $\alpha+\beta$ of
the two beta distribution parameters to have a given margin of error
if $\alpha=\beta$.'' However, \citet{wolk2023star} used an Approval
Voting poll for all voting methods, whereas here we allow the viability-aware
strategies for different voting methods, so we replace each candidate's
percent approval with their support in the poll. In polls for STAR
elections, this is expressed as a fraction of the maximum possible
score rather than as the candidate's average score.}. We also define the expected value of the election $EV=\sum_{j}p_{j}u_{j}$,
where $p_{j}$ is the $j$th candidate's estimated probability of
winning and $u_{j}$ is the utility assigned to that candidate. Note
that this definition implicitly depends on the type of poll used.

The viability-aware strategies are as follows; the strategies for
Plurality, Approval, IRV, and STAR are the same as in \citet{wolk2023star}
except that they use different polls
\begin{itemize}
\item Plurality: Vote for $\mathrm{\textrm{argmax}}_{j}\left(p_{j}\left(u_{j}-EV\right)\right)$.
\item Plurality Top 2: Vote for $\textrm{argma\ensuremath{x_{j}}}\left(p_{j}\left(u_{j}-EV\right)\right)$.
\item Approval: Vote for all candidates with $u_{j}\geq EV$.
\item Approval Top 2: Vote for all candidates with $u_{j}\geq EV$.
\item Instant Runoff Voting: Rank all candidates with $u_{j}\geq EV$ in
decreasing order of $p_{j}u_{j}$. Rank the other candidates sincerely.
\item STAR Voting: Let $f_{j}=\sum_{k}p_{j}p_{k}\left(1-p_{j}-p_{k}\right)\left(u_{j}-u_{k}\right)$
and $r_{jk}=p_{j}p_{k}\left(u_{j}-u_{k}\right)$. The factor $p_{j}p_{k}\left(1-p_{j}-p_{k}\right)$
is an estimator for how likely candidates $j$ and $k$ are to be
in a two-way tie for second place in the scoring phase and $p_{j}p_{k}$
is an estimator for how likely candidates $j$ and $k$ are to face
off in the runoff. Let $s_{j}$ be the score assigned to the $j$th
candidate and let $\textrm{sign}(x)$ be 1 if $x$ is positive, -1
if $x$ is negative, and 0 if $x=0$. Cast the ballot that maximizes\footnote{Strictly speaking, the code approximates the maximum by using an algorithm
that only considers altering the score given to a single candidate
in each optimization step.} $\sum_{j}\left(q\cdot s_{j}f_{j}+\sum_{k}\textrm{sign}\left(s_{j}-s_{k}\right)r_{jk}\right)$,
where $q=0.1$ (see Appendix \ref{sec:Appendix-1:ESIF}). Here, the
first term corresponds to the voter's strategic incentive to give
high scores to the candidates they want to advance to the runoff and
low scores to the ones they don't, and results in strategic exaggeration.
The second term corresponds to the incentive to give candidates (especially
the frontrunners) different scores to have one's vote be impactful
in the runoff.
\item Minimax: Vote sincerely\footnote{While a viability-aware strategy was included for Minimax in \citet{wolk2023star},
this strategy was not incentivized.}.
\end{itemize}
The viability-aware strategies for Plurality and Approval use polls
in which voters vote sincerely using the voting method in question.
For the poll used by the viability-aware STAR strategy, voters cast
sincere ballots but the runoff stage is skipped for tabulating the
poll results. For Plurality Top 2, Approval Top 2, and IRV, the polls
employ Approval Voting such that the voters rescale the candidates'
utilities such that their favorite has 1 utility and the average candidate
is assigned 0 utility, and then vote for all candidates with a utility
of at least 0.1\footnote{In Appendix \ref{sec:Appendix-1:ESIF} we show this number is approximately
strategically optimal for each of these three voting methods.}. The difference in polls means that strategies with identical descriptions
above (such as Plurality and Plurality Top 2) involve different voter
behavior since $p_{j}$ and $EV$ are determined by different polling
data.

All polls show the \% support for each candidate, and for all polls,
we add noise for each candidate that is normally distributed with
a standard deviation of 10\%\footnote{The apparent support in polls is then clipped such that a candidate
cannot appear to have negative support or to be supported by over
100\% of voters.}. For Plurality, Approval, and STAR, this amount is used directly
in the estimations of each candidate's probability of winning. The
Plurality Top 2, Approval Top 2, and IRV strategies assume twice as
much polling uncertainty to account for the extra differences between
the poll and the election.

\begin{figure}

\includegraphics[scale=0.7]{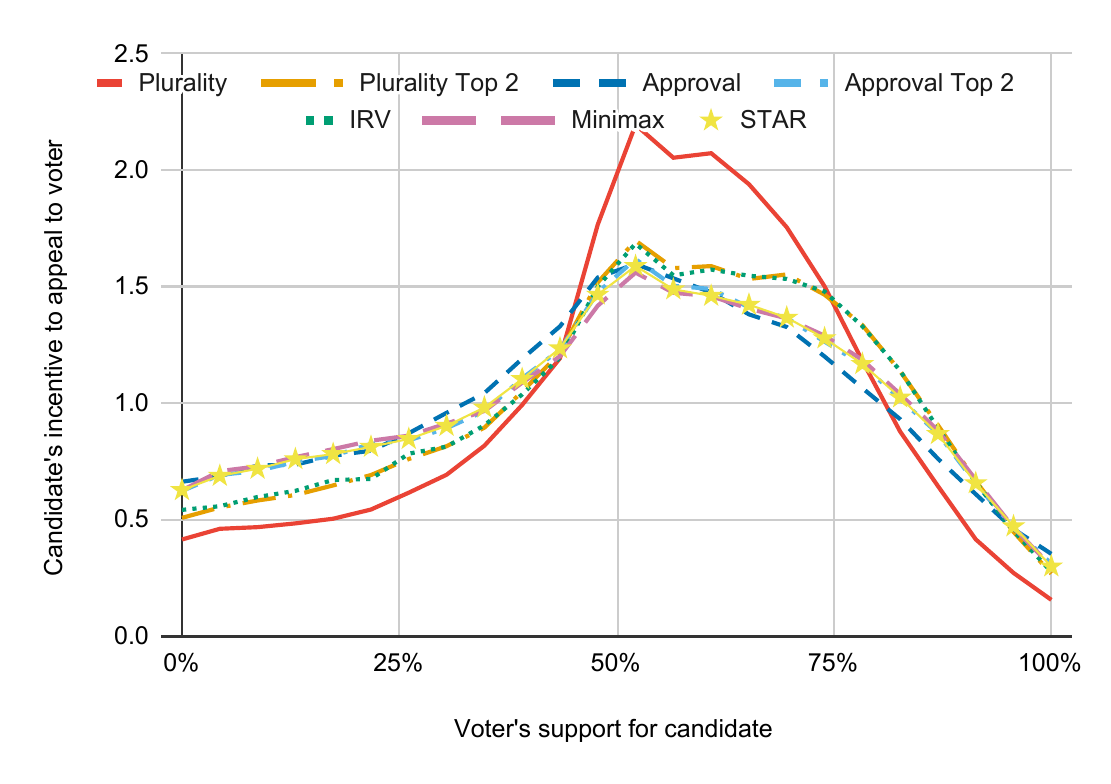}

\caption{\label{fig:Viability-aware-5}Viability-aware voters, 5 candidates,
50,000 iterations}
\end{figure}

Figure \ref{fig:Viability-aware-5} shows each CID with viability-aware
voters. As is the case with sincere voters, Plurality has the highest
peak and the weakest incentives to appeal to strongly opposed voters\footnote{This chart most likely understates the importance under Plurality
of appealing to one's base since our definition of CID does not account
for appealing to a group to perform well in the polls. For Plurality,
this incentive is heavily skewed toward relatively supportive voters.}. Plurality Top 2 and IRV have incentives to appeal to opposing voters
that lie between Plurality and the other methods. However, the differences
between voting methods with viability-aware voters are much smaller
than these differences are with sincere voters. In particular, the
incentive to appeal to opposing voters is dramatically stronger under
Plurality, Plurality Top 2, and IRV. This finding is in line with
Horowitz's \citeyearpar{horowitz2004alternative} argument that strategic
voting can help elect moderates under IRV.

\begin{figure}
\includegraphics[scale=0.7]{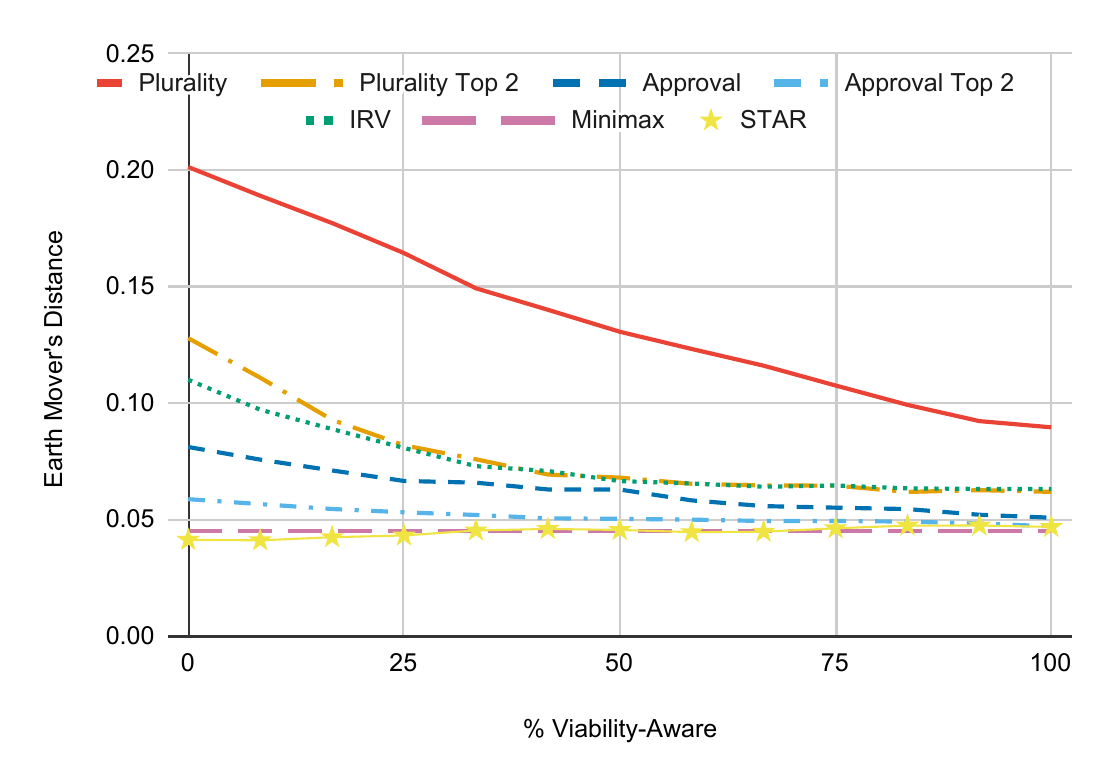}\caption{\label{fig:EMU-va}EMU based on \% viability-aware voters, 5 candidates,
10,000 iterations}

\end{figure}

Figure \ref{fig:EMU-va} shows that, for all methods except Plurality
that benefit from strategic voting, virtually all of these benefits
come from having the first 60\% of the electorate vote in a viability-aware
manner. While a proper investigation of vote psychology lies well
outside the scope of this paper, some voting methods may cause more
voters to consider candidate viability than others. For example, if
voters will vote strategically more often in simple voting methods
with easily-understandable strategic incentives\footnote{See, e.g., \citet{complexity_strategy_rates}},
we should expect more viability-aware voting under Plurality and Approval
than under IRV.

\section{Results for multi-winner elections\label{sec:Results-for-multi-winner}}

We consider the following multi-winner voting methods:
\begin{itemize}
\item Block Plurality: Voters vote for as many candidates as there are winners.
The candidates with the most votes win.
\item Block Approval: Voters vote for any number of candidates.\footnote{We use the same sincere strategy as for single-winner Approval Voting.}
The candidates with the most votes win.
\item Single Nontransferable Vote (SNTV): Each voter votes for only one
candidate. The candidates with the most votes win.
\item Single Transferable Vote (STV): Voters rank the candidates. At the
beginning of tabulation, each ballot is counted as one vote for its
highest-ranked candidate. In each round, any candidates who have a
Droop quota\footnote{In a $w$-winner election with $n$ voters, the Droop quota is $\mathrm{floor}\left(n/\left(w+1\right)\right)+1$. }
of votes are elected, and their surplus votes are transferred to the
highest remaining choices on their supporters' ballots.\footnote{We use the Weighted Inclusive Gregory method for surplus handling.}
If no candidate receives a Droop quota of votes, the candidate with
the fewest votes is eliminated and their votes are transferred as
in IRV.
\item STV-Minimax: Uses the same ranked ballots and algorithm for electing
candidates and performing transfers as STV, but with a different algorithm
for eliminating candidates. Under STV-Minimax, the candidate whose
greatest margin of head-to-head victory against other remaining candidates
is the smallest is eliminated instead of the candidate with the fewest
votes. (This is the Condorcet loser, if there is one.) Margins of
head-to-head victory are determined at the beginning of tabulation.
\end{itemize}
Here we consider elections with four winners and 20 candidates (the
same ratio as for five-candidate single-winner elections). On account
of the complexity of modeling strategic voting in multi-winner elections,
we only consider sincere voters.

\begin{figure}

\includegraphics[scale=0.7]{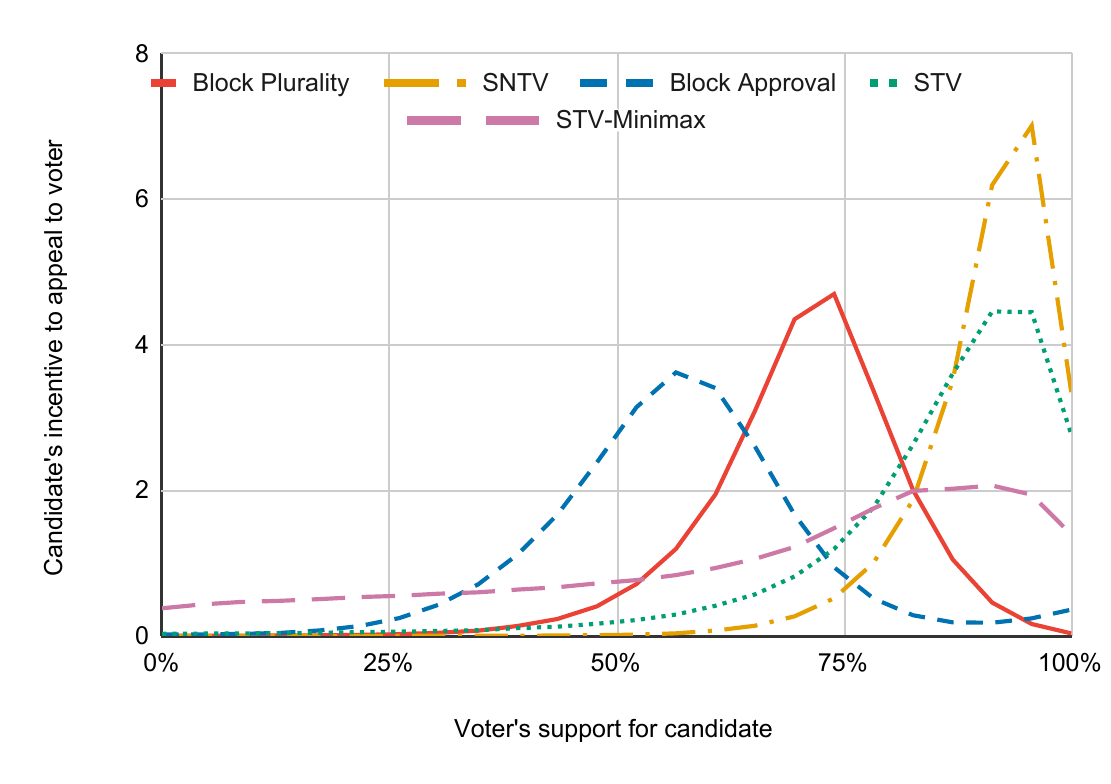}\caption{\label{fig:cid_mw}Sincere voters, 20 candidates, 4 winners, 50,000
iterations}

\end{figure}

Figure \ref{fig:cid_mw} shows that none of these voting methods have
an especially even CID. Block Approval and Block Plurality have similar
CIDs to their single-winner versions. SNTV, STV, and STV-Minimax all
have peaks well to the right, suggesting that incentives to prioritize
appealing to supportive voters over opposing voting are a natural
consequence of proportional representation. However, there are major
differences between these methods. With SNTV, even the incentive to
appeal to the median voter is practically nonexistent; SNTV incentivizes
strong appeal to what can be a very narrow segment of the electorate.
STV incentivizes candidates to care about a somewhat broader range
of voters, but it has a more lopsided CID than any single-winner voting
method in Figure \ref{fig:hon5c}; incentives for appealing to opposing
voters in STV seem negligible. On the other hand, STV-Minimax shows
that it is possible for a proportional voting method to provide a
meaningful incentive to appeal even to strongly opposed voters; the
Candidate Incentive at $x=0$ is 38\%, compared to 3\% for conventional
STV. While STV-Minimax's CID is skewed to the right-hand side of the
chart, it is moderately more balanced than IRV's.

We do not include results for party list methods. CID can be generalized
to describe the electoral incentives faced by parties in the form
of Party Incentive Distributions (see Appendix \ref{sec:Appendix-generalizations}).
However, providing apples-to-apples comparisons between party list
methods and other methods is problematic. This is partly because party
list methods are often used with higher district magnitudes than the
other methods we consider, and partly because the results would be
heavily influenced by the approach used to generate parties. The incentives
faced by a party are closely dependent on its size; a party expected
to win a fraction $z$ of all votes is incentivized to appeal mostly
to voters near $x=1-z$, so a Party Incentive Distribution closely
reflects the distribution of party sizes. We can also consider Candidate
Incentive Distributions for open list methods and Mixed-Member Proportional
Representation (MMP), but these are relatively uninteresting; depending
on the modeling assumptions, an open list method can behave quite
similarly to SNTV, and for candidates running for constituency seats
under MMP, the CID is simply that of the single-winner method used
for the individual constituency.

\section{Conclusion}

Creating incentives for compromise and cooperation is at the heart
of many political reforms, both in newly-fledged democracies in developing
countries and in the United States. Insofar as electoral incentives
shape the political landscape, selecting a voting method that incentivizes
candidates to appeal to a wide range of voters is one of the most
promising approaches to promoting political compromise, stopping ethnic
violence, and preventing civil wars.

CID, by itself, does not tell us whether the centripetal approach
can be effective in general. What it can tell us is when a voting
method fails to provide meaningful centripetal incentives altogether
and which voting methods present stronger centripetal incentives than
others. Additionally, CID can be used to make predictions about the
relative strengths the centripetal effects of voting methods in different
societies by considering how the composition of a society determines
which incentives would have a centripetal effect. Here we consider
the predictions CID makes in four societies:

In Papua New Guinea, where in the 1997 national election (conducted
using Plurality) ``over half of all seats were won with less than
20 percent of the vote'' \citep{reilly2004}, even incentives to
appeal to voters at $x=60\%$ are likely to serve a centripetal role.
While simulations optimized for reflecting the politics of Papua New
Guinea would have considerably more candidates, the ten-candidate
simulations in Figure \ref{fig:hon10c} show that IRV's CID is well
above 1 (unlike Plurality's CID) when $x$ is between 50\% and 60\%,
which suggests that it provides powerful centripetal incentives in
the context of Papua New Guinea. Empirical findings bear out this
prediction; \citet{reilly2004} writes that Papua New Guinea ``offers
perhaps the most compelling case for the use of preferential voting
as a means of conflict management in ethnically-diverse societies.''

In Fiji, CID suggests that IRV provided much weaker centripetal incentives
than it did in Papua New Guinea, but does not offer clear predictions
beyond this. Given that the ethnic divide in Fiji is primarily about
two competing ethnic groups, we should not usually expect voters at
$x>50\%$ to provide centripetal incentives, and Figure \ref{fig:hon5c}
shows that appealing more to voters at $x>50\%$ is nearly three times
as valuable, on average, as appealing more to voters at $x<50\%$.
The incentive to appeal to less supportive voters is meaningful, but
not necessarily stronger than the centrifugal incentives present in
Fijian society. It should also be noted that using CID to analyze
incentives in the Fijian context is complicated by Fiji's ``above
the line'' voting in which a voter could select a single party instead
of ranking candidates, and their vote would be transferred according
to that party's predetermined rankings \citep{arms2006Fiji}. This
is a significant deviation from ``conventional'' IRV that is not
accounted for in our simulations.

In Northern Ireland, our findings suggest that centripetal electoral
incentives did not play a significant role in leading to a lasting
peace because the centripetal incentives of STV are so weak as to
be negligible; appealing more to a voter at $x=95\%$ is approximately
23 times as valuable as appealing the same amount more to a voter
at $x=50\%$. This is consistent with the observations of \citet{jarrett2017peace}
that ``parties do not, on the whole, campaign with the specific intention
of attracting transfers'' and that ``parties have given little thought
to any potential role for inter-bloc transfers'' (between unionists
and nationalists).

The United States, unlike the above examples, is divided primarily
along partisan lines rather than ethnic lines \citep{Reilly_established_democracies},
and the split into two distinct groups is more reminiscent of Fiji
and Northern Ireland than of Papua New Guinea. A major difficulty
with centripetal reforms in the US is that self-described ``moderate''
voters tend to be the least politically engaged and not especially
moderate. As \citet{drutman_primary} writes, ``Many self-identified
independents and moderates are more politically extreme than partisans.
They are also less likely to reward compromise than registered partisans,
contrary to conventional wisdom.'' Thus, we should not expect large
centripetal effects from incentives to appeal to voters near $x=50\%$
in the US. Since IRV's Candidate Incentive is approximately 1/3 for
strongly opposed voters for whom an incentive to appeal toward should
be centripetal, our findings suggest that, while IRV yields more balanced
incentives than Plurality, the effects of switching to it are unlikely
to be dramatic. There may be cause for greater optimism with the recent
adoption of IRV in Alaska, however; \citet{reilly2023alaska} write
that Alaska satisfies the conditions for IRV to be effective ``perhaps
more than any other state'' with more genuinely moderate voters,
so centripetal incentives may come from a larger fraction of the electorate
there than elsewhere in the US.

The US is also a testing ground for methods other than IRV. Approval
Voting and Block Approval were adopted in Fargo, North Datoka in 2018
\citep{fargo_adoption}, and Approval Top 2 was adopted in St. Louis
in 2020 \citep{STL_adoption}. Eugene, Oregon will vote in May of
2024 on whether to adopt STAR Voting \citep{Eugene_initiative}. While
it can be difficult to assess the centripetal tendencies of voting
methods solely from municipal elections, our findings suggest that
jurisdictions that adopt these voting methods (especially Approval
Top 2 and STAR Voting) will see greater centripetal effects than those
that adopt IRV.

By asking which voters different voting methods incentivize candidates
to appeal to, we can evaluate a long-held assumption about centripetalism.
Regarding the conditions for a centripetalist approach to be effective,
\citet{reilly2004} writes:
\begin{quote}
A key facilitating condition appears to be the presence of a core
group of moderates, both amongst the political leadership and in the
electorate at large. Centripetal strategies for conflict management
assume that there is sufficient moderate sentiment within a community
for cross-ethnic voting to be possible.
\end{quote}
The presence of moderate voters is certainly helpful, especially under
voting methods like IRV where the incentive to appeal to them is dramatically
greater than the incentive to appeal to opposing extremists. However,
our finding that several voting methods provide strong incentives
to appeal to even the most staunchly opposed voters suggests that
this moderate core may not be strictly necessary. We can identify
two types of centripetal incentives. The first of these, which has
been the main focus of prior research, is the incentive to appeal
to moderate voters who are close to the political center. The second
is the incentive to appeal to opposing extremist voters, as well as
opposing ``moderate'' voters who would insist on giving all members
of their own ethnic group greater support than the moderate candidates
of other ethnic groups\footnote{In the US context we would replace ``ethnic group'' with ``political
party''.}. IRV, Plurality Top 2, and especially Approval Voting are all reasonably
effective at creating the first type of centripetal incentive, but
do not provide a particularly large centripetal incentive of the second
type. Condorcet methods, STAR Voting, Approval Top Two, and (to a
lesser extent) STV-Minimax provide significant incentives of both
types and therefore may be able to exhibit centripetal tendencies
even in societies where moderate voters are a small minority.

Prior research has focused on the claim that IRV and STV provide centripetal
incentives that are lacking in Plurality. Our findings support this
claim in the case of IRV but contest it in the case of STV. The main
novel contribution of this paper is in evaluating electoral incentives
under additional voting methods. We have found that Condorcet Methods,
STAR Voting, and Approval Top Two give candidates comparable incentives
to appeal to all voters, suggesting that centripetalist reformers
should give these methods serious consideration. These findings hold
under a wide range of voter behavior.

There are many possible directions for future research. Party Incentive
Distributions could be used to compare STV with ``above-the-line''
voting to party list systems. Custom voter models could be designed
to reflect a particular society in which centripetal reforms are being
considered. Another research avenue is to use a spatial voter model
with clusters at fixed locations and consider the relative strengths
of the incentives for a candidate at a given point in this spatial
model to appeal to voters at different points; this would make it
possible to distinguish between the incentives presented to moderate
and to extremist candidates.

\specialsection*{Code Availability}

The source code is available at \href{https://github.com/ragconsumer/VMES}{https://github.com/ragconsumer/VMES}.

\specialsection*{Acknowledgements}

Thanks to Jeanne Clelland for reading early drafts and making many
helpful suggestions, and to Arend Peter Castelein for help with the
charts.

\bibliographystyle{cas-model2-names}
\bibliography{citations}

\appendix

\section{Party Incentive Distributions\label{sec:Appendix-generalizations}}

For defining Party Incentive Distributions, we use the same notation
as for defining Candidate Incentive Distributions aside from the following
changes:
\begin{itemize}
\item $m$ refers to the number of parties instead of the number of candidates.
\item $f$ is a function that, instead of mapping vectors of ballots to
sets of winning candidates, maps vectors of ballots to $\mathbb{Z}^{m}$.
When \textbf{$\vec{b}$} is a vector of ballots, $f\left(\vec{b}\right)_{i}$
is the number of seats won by the $i$th party.
\item $i$ is an index over parties. In the subsequent definitions we take
it to be a uniformly distributed.
\end{itemize}
\begin{defn}
The \emph{Unnormalized Unweighted Party Incentive} is
\end{defn}

\begin{align*}
 & UUPI\left(f,\vec{t},n,m,H,\epsilon,S,p,K,k\right)=\\
 & \quad\mathbb{E}\left[f\left(\vec{t}\left(G,p(G,\xi)\right)\right)_{i}-f\left(\vec{t}\left(S_{c;G}^{T}\left(S_{c;G}G-\epsilon^{k,c}\right),p(G,\xi)\right)\right)_{i}\right]\\
 & +\mathbb{E}\left[f\left(\vec{t}\left(S_{c;G}^{T}\left(S_{c;G}G+\epsilon^{k,c}\right),p(G,\xi)\right)\right)_{i}-f\left(\vec{t}\left(G,p(G,\xi)\right)\right)_{i}\right]
\end{align*}
There is one issue with this definition: it considers the incentives
faced by small parties and large parties to be equally important,
so that the incentives governing a seat held by a large party are
less influential on the UUPI than the incentives governing a seat
held by a small party. We can address this by weighting the contribution
of each party by the number of seats it wins, average across the perturbed
and unperturbed cases.
\begin{defn}
Write $\vec{b}:=\vec{t}\left(G,p(G,\xi)\right)$, $\vec{b^{+}}:=\vec{t}\left(S_{c;G}^{T}\left(S_{c;G}G+\epsilon^{k,c}\right),p(G,\xi)\right)$,
and $\vec{b^{-}}:=\vec{t}\left(S_{c;G}^{T}\left(S_{c;G}G-\epsilon^{k,c}\right),p(G,\xi)\right)$.
The \emph{Unnormalized Weighted Party Incentive is}
\begin{align*}
 & UWPI\left(f,\vec{t},n,m,H,\epsilon,S,p,K,k\right)=\\
 & \quad\mathbb{E}\left[\left(f\left(\vec{b}\right)_{i}-f\left(\vec{b^{-}}\right)_{i}\right)\cdot\frac{\left(f\left(\vec{b}\right)_{i}+f\left(\vec{b^{-}}\right)_{i}\right)}{2}\right]\\
 & +\mathbb{E}\left[\left(f\left(\vec{b^{+}}\right)_{i}-f\left(\vec{b}\right)_{i}\right)\cdot\frac{\left(f\left(\vec{b}\right)_{i}+f\left(\vec{b^{+}}\right)_{i}\right)}{2}\right]
\end{align*}
\end{defn}

From here, the normalization to create a weighted (or unweighted)
Party Incentive Distrbution is exactly the same as for CID.

\section{Assessing strategic voting\label{sec:Appendix-1:ESIF}}

To evaluate the extent to which a strategy is incentivized, we define
Expected Strategic Influence Factors (ESIF).
\begin{defn}
\label{def:ESIF}Let $w$ be the winner\footnote{ESIF can also be generalized to multi-winner voting methods by replacing
the $i$th voter's utility of the single winning candidate with some
function of the $i$th voter's utilities of all the winning candidates,
such as the mean, median, or maximum.} of an election if all voters vote in accordance with the strategy
vector $\vec{t}$, i.e. $w=f\left(\vec{t}\left(G,p\left(G,\xi\right)\right)\right)$.
Similarly, let $w_{i}^{s}$ be the winner if instead the $i$th voter
uses the strategy $s$ while all other voters vote in accordance with
$\vec{t}$, and let $w_{i}^{a}$ be the winner if the $i$th voter
abstains while all other voters vote in accordance with $\vec{t}$.
The ESIF of strategy $s$ relative to $\vec{t}$ for the $i$th voter
is then

\[
ESIF\left(s\vec{;t},i\right)=\frac{E\left[G_{iw_{i}^{s}}-G_{iw_{i}^{a}}\right]}{E\left[G_{iw}-G_{iw_{i}^{a}}\right]}
\]
\end{defn}

The expected values in Definition \ref{def:ESIF} are with regard
to the random matrix $G$ (distributed according to a voter model
$H$) and polling noise $\xi$. Note that ESIF is also a function
of the voting method, number of candidates, etc. Intuitively, $ESIF\left(s\vec{;t},i\right)$
says how many times more influential the $i$th voter is if they use
$s$ instead of $t_{i}$; an ESIF of 1 means that $s$ is exactly
as effective as $t_{i}$, an ESIF of 1.5 means that a voter wields
50\% more influence if they use $s$ instead of $t_{i}$, and an ESIF
of 0 means that a vote according to $s$ fails to advance the voter's
interests at all (in expectation).

To determine our ``sincere'' strategy for Approval and Approval
Top 2 we use ESIF to find an evolutionarily stable strategy of the
form, ``rescale one's utilities to a scale such that one's favorite
candidate is assigned a utility of 1 and the average utility of the
candidates is 0, and to vote for all candidates whose utility is at
least $z$''.

\begin{figure}
\includegraphics[scale=0.7]{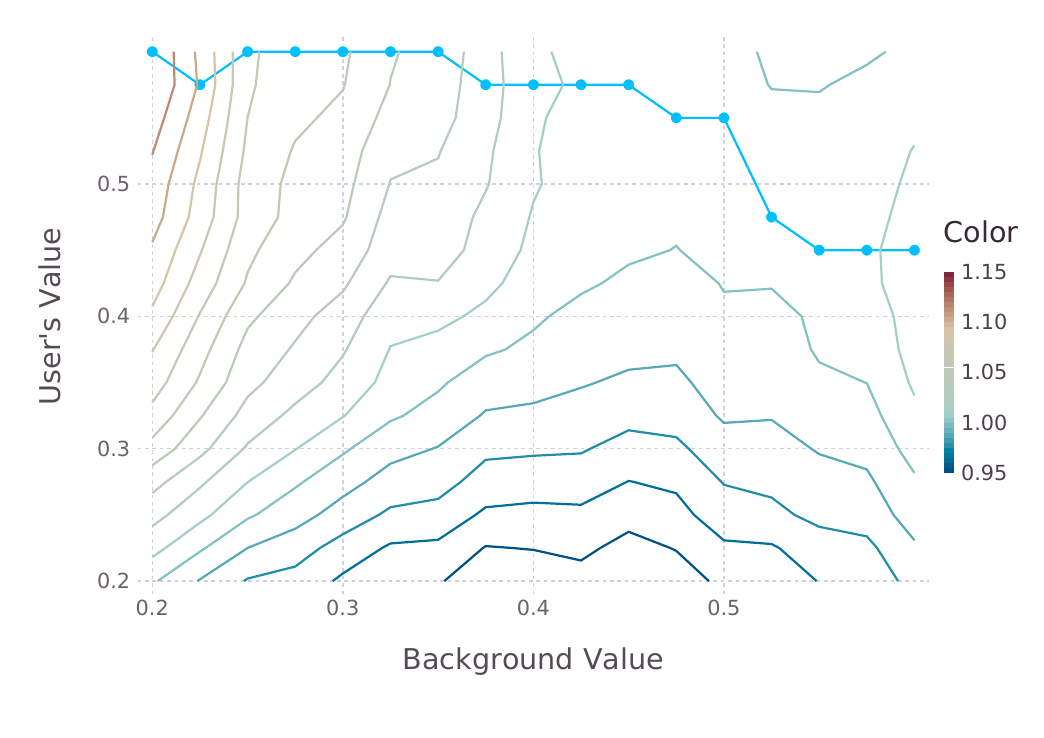}\caption{\label{fig:ESIF-for-Approval}ESIF for Approval Voting, 5 candidates,
31 voters, 15000 iterations}
\end{figure}

\begin{figure}

\includegraphics[scale=0.7]{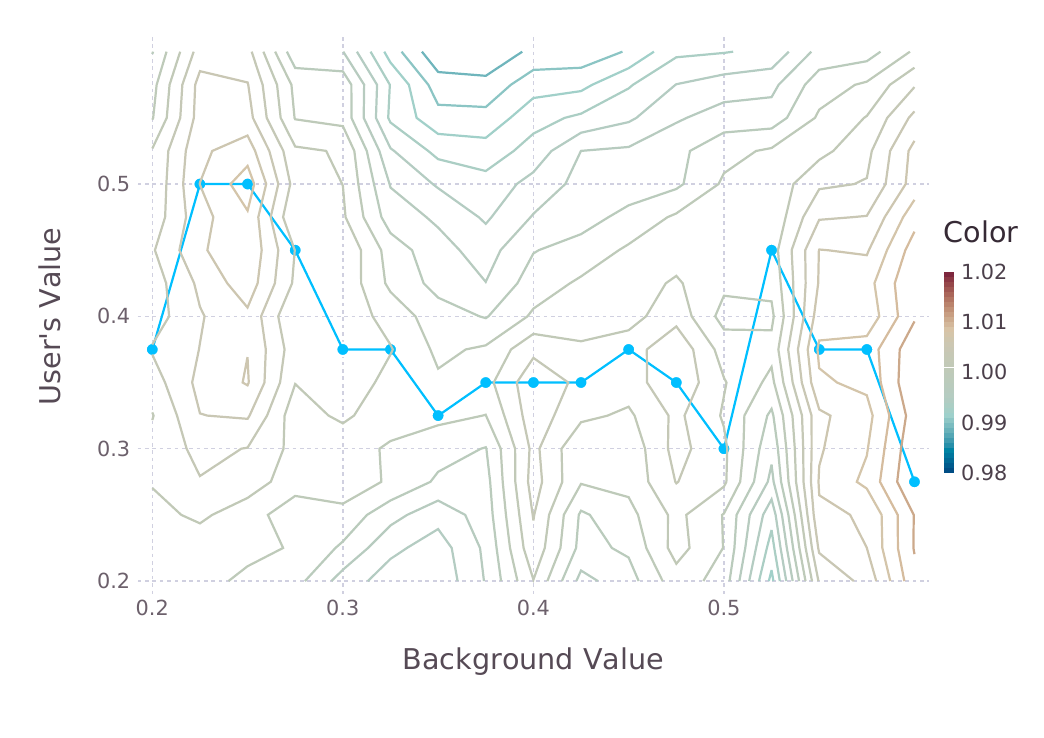}\caption{\label{fig:ESIF-Approval2}ESIF for Approval Top 2, 5 candidates,
31 voters, 15000 iterations}

\end{figure}

Figures \ref{fig:ESIF-for-Approval} and \ref{fig:ESIF-Approval2}
use contour plots to show the ESIFs of strategies of the above form.
The y-axis gives the value of $z$ used by the voter in question,
and the x-axis gives the value of $z$ used by the rest of the electorate.
The blue line shows the most strongly incentivized strategy; wherever
this line intersects the line $y=x$ is an evolutionarily stable strategy.
We find that these occur at approximately 0.5 for Approval Voting
and at 0.35 for Approval Top 2. However, ESIF simulations tend to
be dominated by noise in the vicinity of the optimal strategy\footnote{This is a consequence of elementary calculus. Assuming ESIF is differentiable,
the derivative of ESIF with respect to the value of $z$ used in $s$
is zero at the optimal value, so as $z$ approaches the optimal value
the signal:noise ratio goes to zero.}, so there is only one significant figure. Taking the average of these
numbers and rounding to a multiple of 0.1 yields the value $z=0.4$.
(While not particularly relevant to CID, it is also interesting to
note the different scales for these plots; as a voter, choosing a
good value of $z$ is much more important under Approval than under
Approval Top 2.)

For the viability-aware strategies, the use of evolutionarily stable
strategies makes less sense because the viability-aware strategies
are not intended as a baseline; for voting methods where voting strategically
is less intuitive, it seems probable that only a small fraction of
voters will behave in a viability-aware manner. Therefore, we optimize
the viability-aware strategies for the case that all other voters
vote sincerely.

\begin{figure}
\includegraphics[scale=0.7]{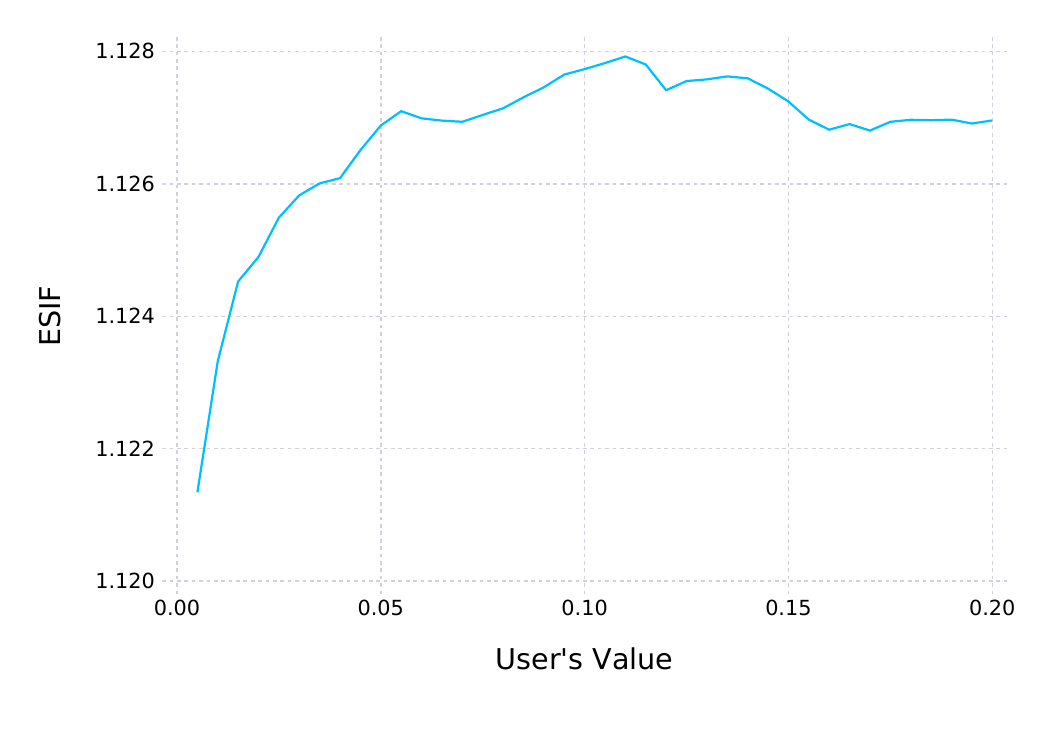}\caption{\label{fig:ESIF-for-STARVA}ESIF for STAR viability-aware's $q$ parameter,
31 voters, 250,000 iterations}

\end{figure}

Figure \ref{fig:ESIF-for-STARVA} shows the ESIF of the viability-aware
STAR strategy for different values of $q$. Note the scale of the
y-axis; $q$ has very little influence on the effectiveness of this
strategy.

\begin{figure}

\includegraphics[scale=0.7]{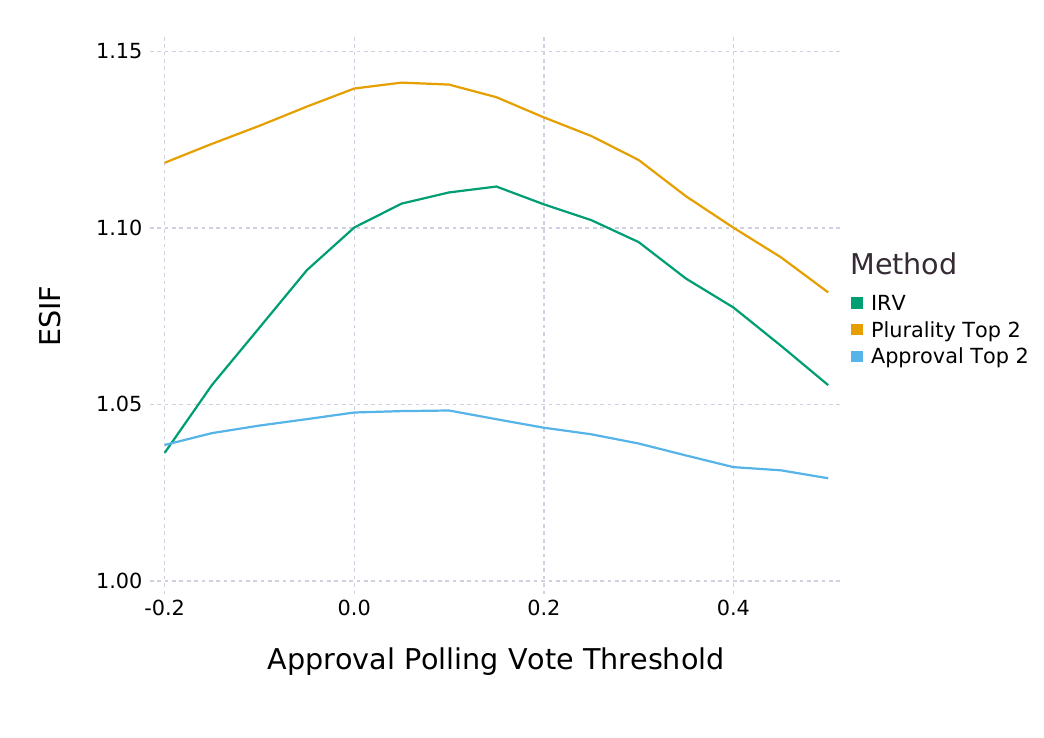}\caption{\label{fig:ESIF-poll-approval-threshold}ESIF for approval thresholds
in polling for viability-aware strategies, 31 voters, 100,000 iterations}

\end{figure}

Figure \ref{fig:ESIF-poll-approval-threshold} shows the ESIF of the
viability-aware strategies for Plurality Top 2, Approval Top 2, and
IRV for different approval thresholds within the Approval Voting poll
used by each of these strategies. For each of these strategies, the
ESIF is maximized at approximately 0.1.

\section{\label{sec:dogmatic-bullet-voting}Results with dogmatic bullet voting}

\begin{figure}
\includegraphics[scale=0.7]{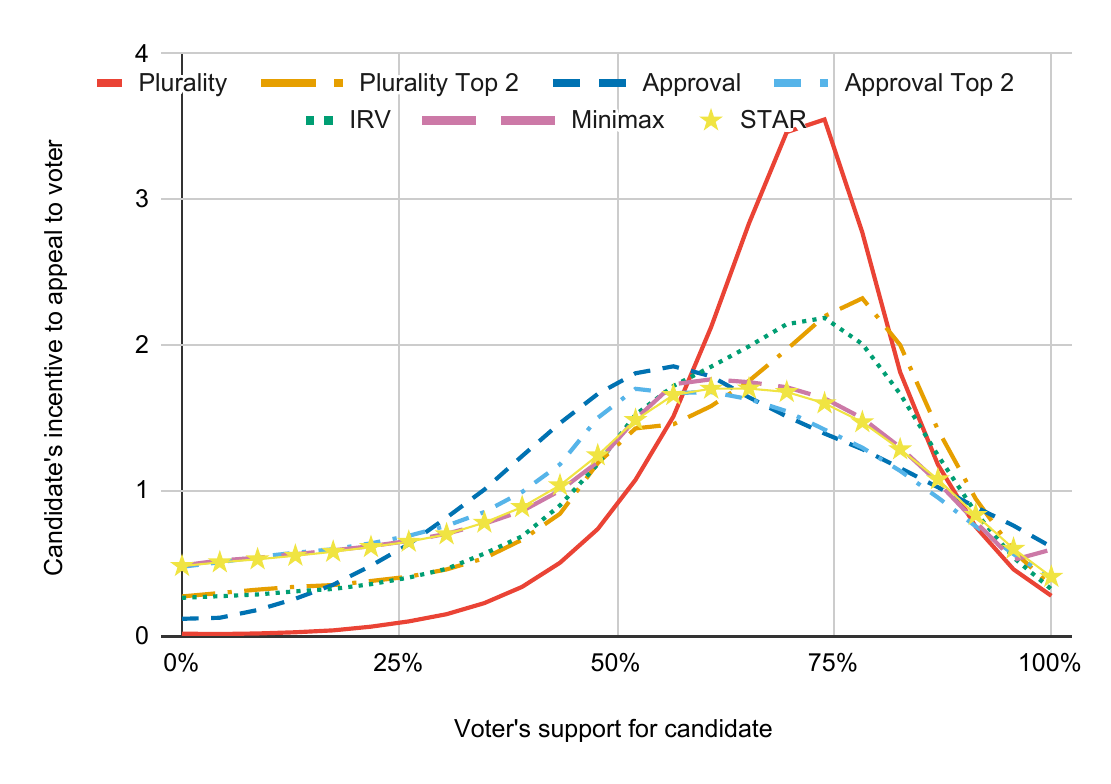}\caption{\label{fig:hon-bullet-CID}71\% sincere voters, 29\% dogmatic bullet
voters, 5 candidates, 100,000 iterations}
\end{figure}

In addition to strategic voters, we should expect other voters to
deviate from the previously-defined ``sincere'' behavior in a manner
that is not strategically incentivized: by ``bullet voting'', or
giving full support to their favorite while giving no support to any
other candidate, regardless of those voters' strategic incentives
or how they feel about candidates other than their favorite. We will
refer to this as ``dogmatic bullet voting'' to distinguish it from
(a) bullet voting which is a good expression of a voter's more nuanced
preferences and (b) bullet voting which is strategically incentivized.
Even though there is no incentive for bullet voting under IRV, \citet{RCVbulletvoting}
has found that only ``A median of 71\% of voters rank multiple candidates''.
To account for this, in Figure \ref{fig:hon-bullet-CID} we have a
randomly chosen 29\% of voters employ dogmatic bullet voting, i.e.
to bullet vote regardless of their preferences. This does not mean
that \emph{only} 29\% of voters will bullet vote; under Approval Voting,
(and, to a much lesser extent, STAR) the sincere strategy that is
employed by the other 71\% of voters can yield bullet voting as well,
but only when it is the best expression of a voter's preferences.
Dogmatic bullet voters will still participate in the runoff of Plurality
Top 2 and Approval Top 2.

The CIDs of IRV and Approval Top 2 are relatively unaffected by this
level of dogmatic bullet voting (Plurality and Plurality Top 2 are,
of course, completely unaffected since everyone bullet votes anyway),
but for Minimax and STAR Voting, the peak moves significantly to the
right as politicians are faced with the Plurality-like incentives
of appealing to the dogmatic bullet voters. Approval Voting is similarly
affected, but since its CID is uncommonly low near Plurality's peak
in the absence of dogmatic bullet voting and its peak is much higher
than that of STAR and Minimax, the change in the shape of Approval's
CID does not significantly affect the location of its peak. The basic
findings are unchanged: STAR, Minimax, and Approval Top 2 have much
more even CIDs than IRV, Plurality, and Plurality Top 2.

\begin{figure}
\includegraphics[scale=0.7]{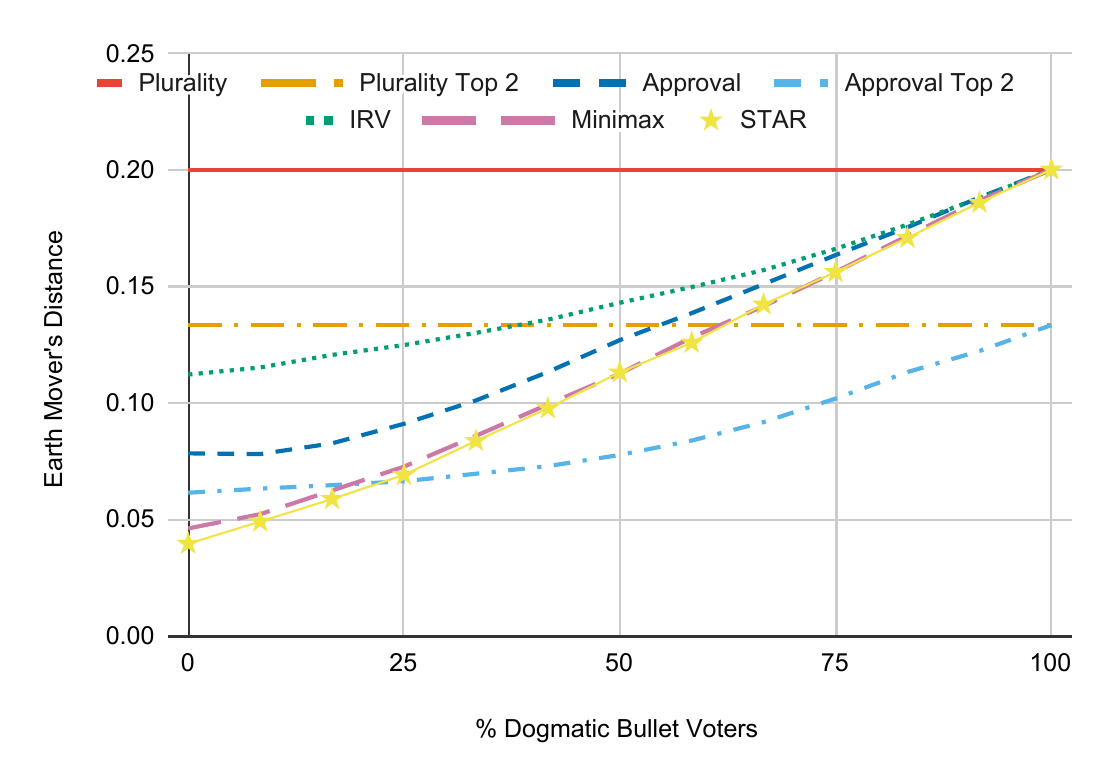}\caption{\label{fig:bullet-EMU}EMU based on \% bullet voters, 5 candidates,
25,000 iterations}
\end{figure}

We can see the effects of higher levels of dogmatic bullet voting
in Figure \ref{fig:bullet-EMU}. When over 25\% of voters bullet vote
regardless of their preferences, Approval Top 2 provides the most
uniform incentives. Aside from that, dogmatic bullet voting has little
effect on which voting methods provide more equitable incentives than
others. While dogmatic bullet voting reduces the magnitude of the
differences, STAR, Score, Approval Top 2, and Minimax all provide
more uniform incentives than IRV regardless of how much of it there
is.

Figure \ref{fig:bullet-EMU} also tells us what happens if dogmatic
bullet voting is more common under some voting methods than others.
Even if nearly half of voters were to dogmatically bullet vote under
STAR, Minimax, and Approval Top 2, all of these methods would yield
more balanced incentives than IRV would without any bullet voting.

\section{Additional results\label{sec:Appendix-2:-Additional-results}}

We start by exploring the effects of using different voter models.
We first consider the extremely simple (and unrealistic) impartial
culture model, in which all voter preferences are uncorrelated, such
that with a large electorate all candidates would be almost equally
popular. The utilities each voter assigns to each candidate are drawn
i.i.d. from a Gaussian distribution.

\begin{figure}

\includegraphics[scale=0.7]{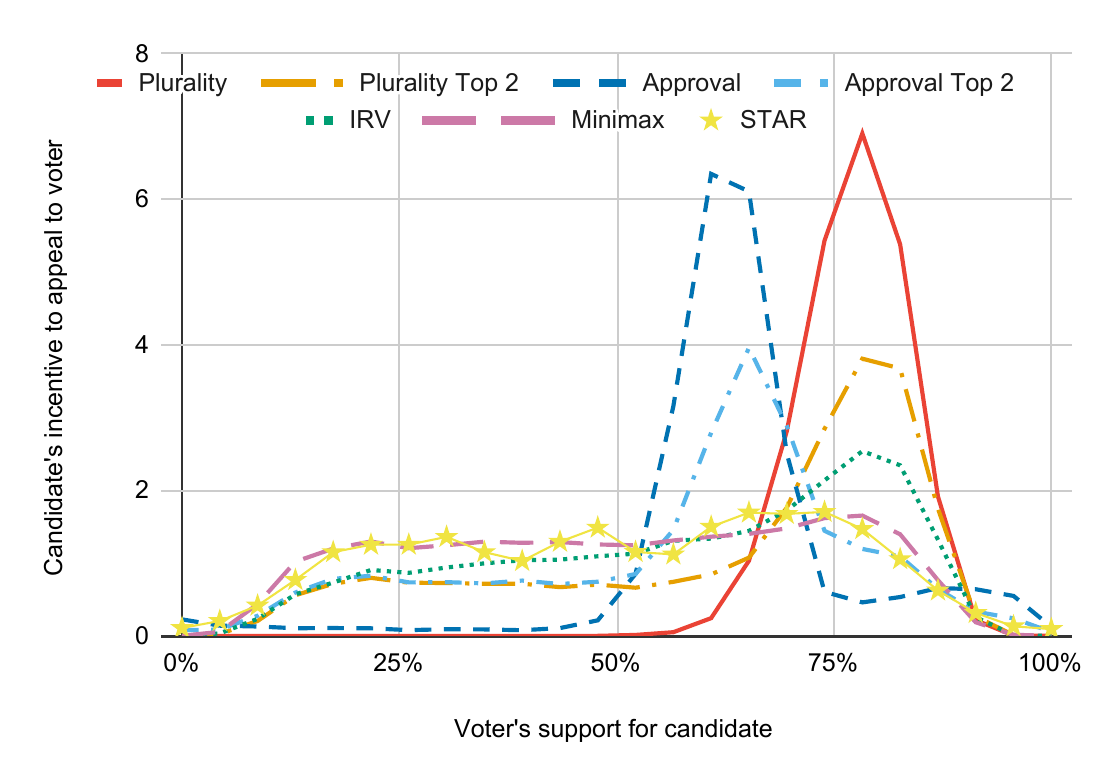}\caption{\label{fig:Impartial-Culture}Impartial culture model, 5 candidates,
sincere voters, 50,000 iterations}

\end{figure}

In Figure \ref{fig:Impartial-Culture} we see that, while it differs
greatly from the corresponding chart for the clustered dimensional
model (Figure \ref{fig:hon5c}), the core findings remain intact:
Plurality has the most lopsided CID, followed by Plurality Top 2.
Under IRV, candidates are significantly more incentivized to appeal
to supportive voters than relatively opposed voters. Minimax and STAR
have the most even CIDs.

However, the differences between figures \ref{fig:Impartial-Culture}
and \ref{fig:hon5c} are worth explaining. Under impartial culture,
the sorting function works much more effectively since complications
that are present in a spatial model (namely the fact that opinions
of different candidates are correlated) are absent. This is responsible
for Approval and Approval Top 2 having peaks that are as sharp as
those of Plurality and Plurality Top 2. It also allows us to see oscillations
in STAR's CID, which are attributable to the fact that it offers six
different levels of support instead of infinitely many, so while it
lacks the ``hard'' thresholds of Approval and Plurality it does
have several ``soft'' thresholds.

Another major difference is that Approval and Approval Top 2 have
peaks that are well to the right of 50\%. This is because the sincere
strategy is miscalibrated for impartial culture. Under a realistic
voter model, the fact that you, a single voter, like a candidate is
(weak) evidence that this candidate is broadly popular, so your favorite
candidate is more likely to win than your least favorite, and you
should take this into account when choosing how many candidates to
vote for. This is not the case under impartial culture, where (with
a large number of voters) it is strategically optimal to vote for
every above-average candidate. For this optimal strategy, the peaks
for Approval and Approval Top 2 would be at 50\%.

Next, we consider simple spatial models in which voters and candidates
are distributed in some ``issue-space'' and voters prefer candidates
who are close to them\footnote{Specifically, voters and candidates are distributed according to a
Gaussian distribution, and a voter's utility for a candidate is minus
the Euclidean distance between them.}. The one-dimensional model can be thought of as the conventional
left-right political axis, for example.

\begin{figure}
\includegraphics[scale=0.7]{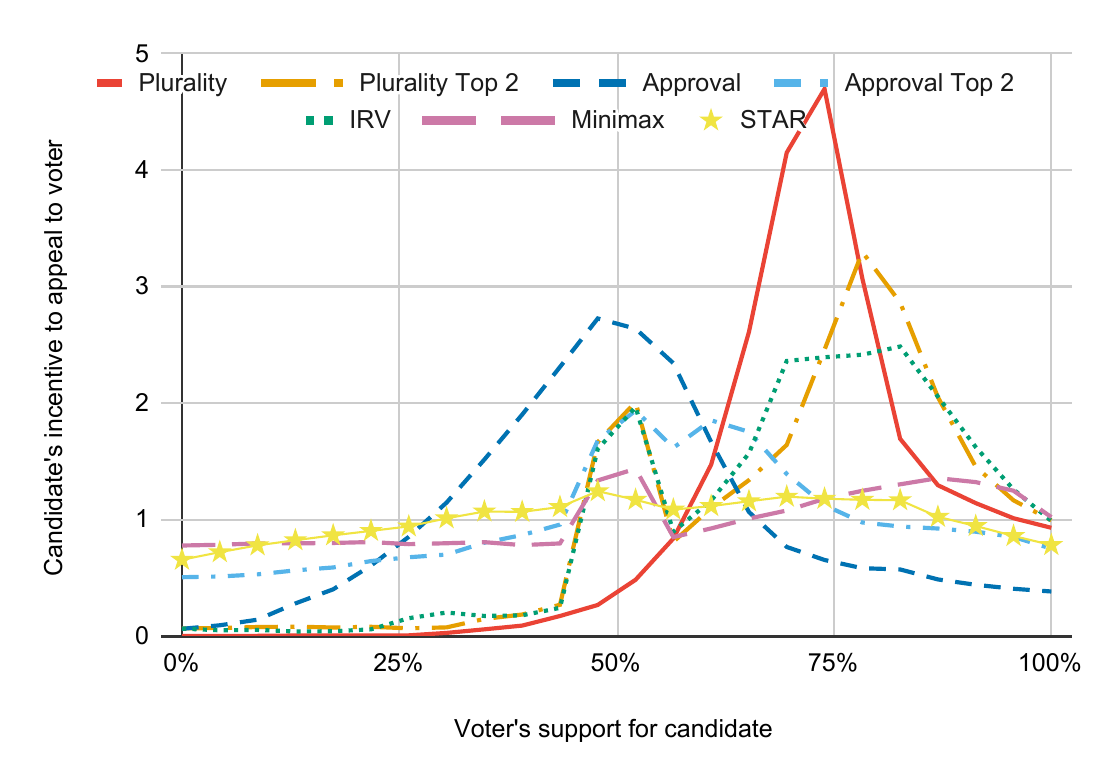}\caption{\label{fig:One-dimensional-spatial-model}One-dimensional spatial
model, 5 candidates, sincere voters, 50,000 iterations}

\end{figure}

In Figure \ref{fig:One-dimensional-spatial-model} we see that the
core findings from the impartial culture and clustered spatial models
continue to hold in the one-dimensional model, but some phenomena
are far more pronounced. Both Plurality Top 2 and IRV have a Candidate
Incentive of well under 0.1 on the leftmost quarter of the graph,
compared to around 0.3 with the clustered spatial model. We also see
that Plurality Top 2 and IRV have a very pronounced secondary peak
at 50\%. These differences are because elections under a 1-dimensional
model have a far more predictable structure. Under IRV, the final
round will usually feature one left-of-center candidate and one right-of-center
candidate. The left-of-center candidate will have no incentive to
appeal to voters who are to the right of the right-of-center candidate
unless the two finalists are very close together ideologically, and
vice versa. The only swing voters will be at the center of the political
spectrum, hence the sharp secondary peak.

\begin{figure}

\includegraphics[scale=0.7]{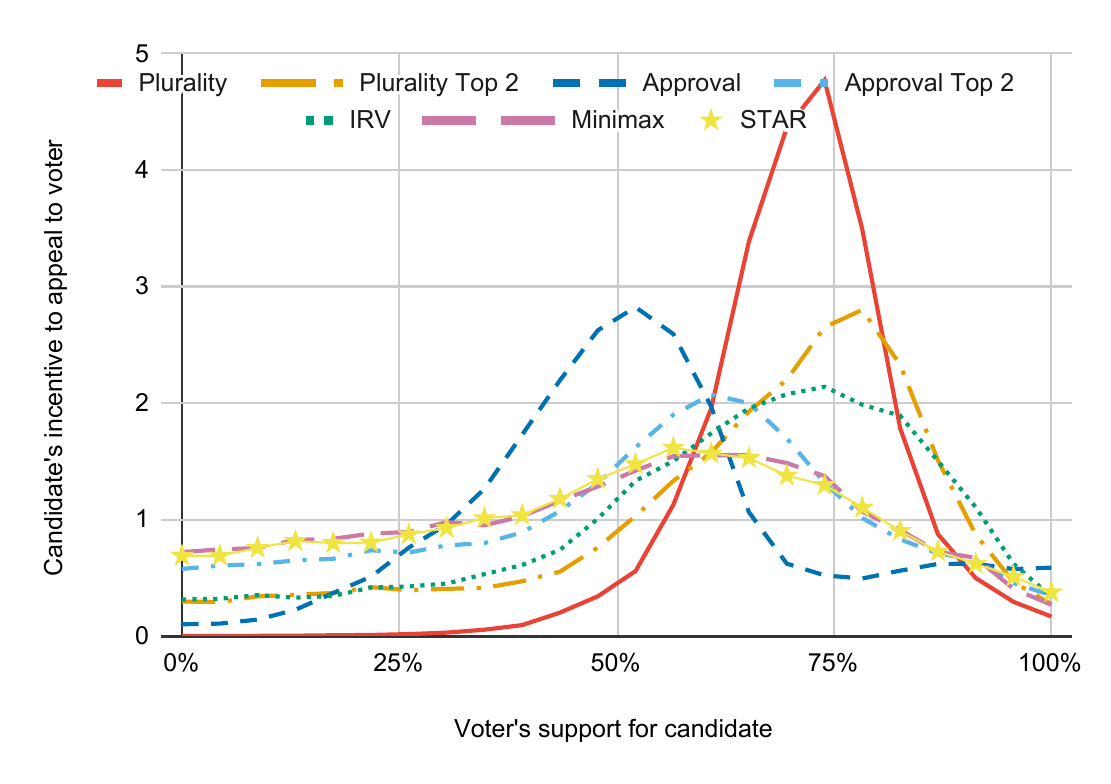}\caption{\label{fig:Two-dimensional-spatial-model}Two-dimensional spatial
model, 5 candidates, sincere voters, 50,000 iterations}

\end{figure}

\begin{figure}
\includegraphics[scale=0.7]{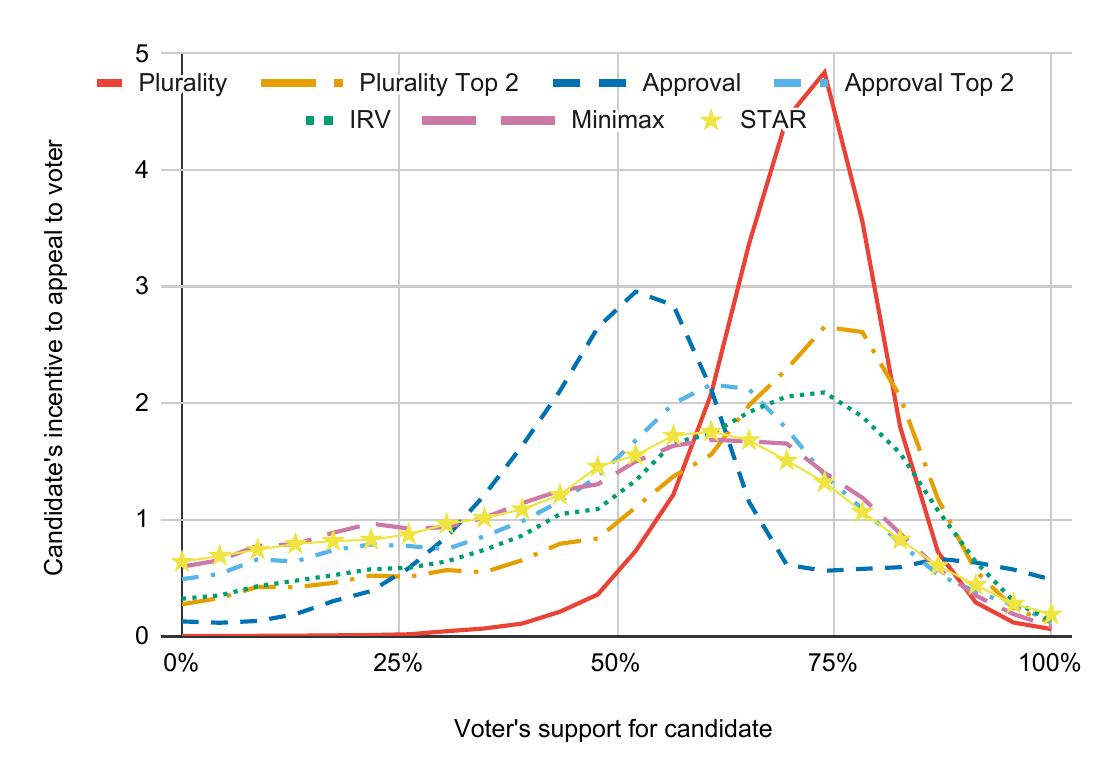}\caption{\label{fig:Three-dimensional-spatial-model}Three-dimensional spatial
model, 5 candidates, sincere voters, 50,000 iterations}

\end{figure}

The graphs for the two- and three-dimensional models in figures \ref{fig:Two-dimensional-spatial-model}
and \ref{fig:Three-dimensional-spatial-model} are very similar to
the one for the clustered spatial model, Figure \ref{fig:hon5c}.
This suggests that while either an extreme lack of structure (Figure
\ref{fig:Impartial-Culture}) or an extremely rigid structure (Figure
\ref{fig:One-dimensional-spatial-model}) can cause oddities, the
exact details of the voter model have a relatively minor influence
on CID.

Next, we turn to the role of the sorting function. In general, sorting
voters from most to least supportive of a candidate is a vaguely-defined
problem that admits several reasonable-sounding solutions, none of
which are entirely satisfactory. We consider three options:
\begin{itemize}
\item Normalized Mean: Rescale each voter's utilities to have mean 0 and
standard deviation 1. Sort them based on their rescaled utility for
the candidate in question. This is the sorting function used in the
main paper.
\item Unnormalized Mean: Shift each voter's utilities to have mean 0. Sort
them by their shifted utility for the candidate in question.
\item Distance from Top: Sort voters based on their utility for the candidate
in question, minus the highest utility assigned to candidates other
than the one in question.
\end{itemize}
\begin{figure}
\includegraphics[scale=0.7]{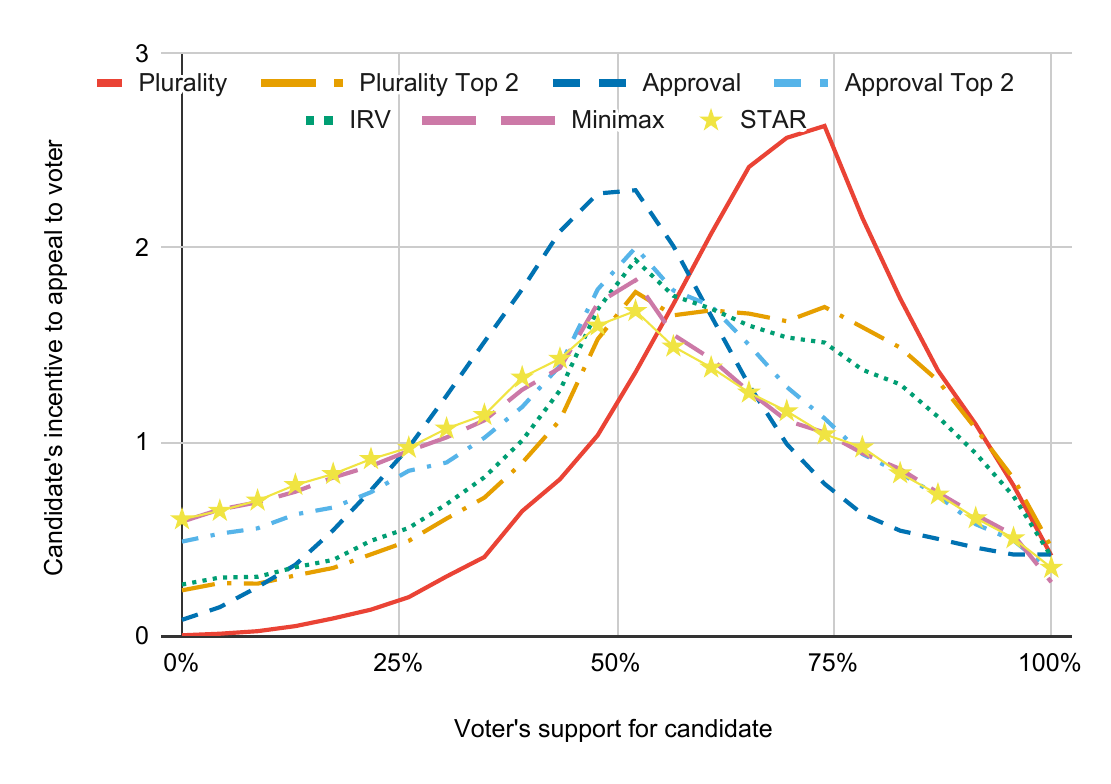}\caption{\label{fig:Unnormalized-sorting-function}Unnormalized Mean, 5 candidates,
sincere voters, 50,000 iterations}

\end{figure}

In figure \ref{fig:Unnormalized-sorting-function} we see far less
of a difference between voting methods than with Normalized Mean,
and only Plurality has a reasonably sharp peak towards the right side
of the graph. However, the left edge of the graph looks extremely
similar for both Normalized Mean and Unnormalized Mean.

\begin{figure}
\includegraphics[scale=0.7]{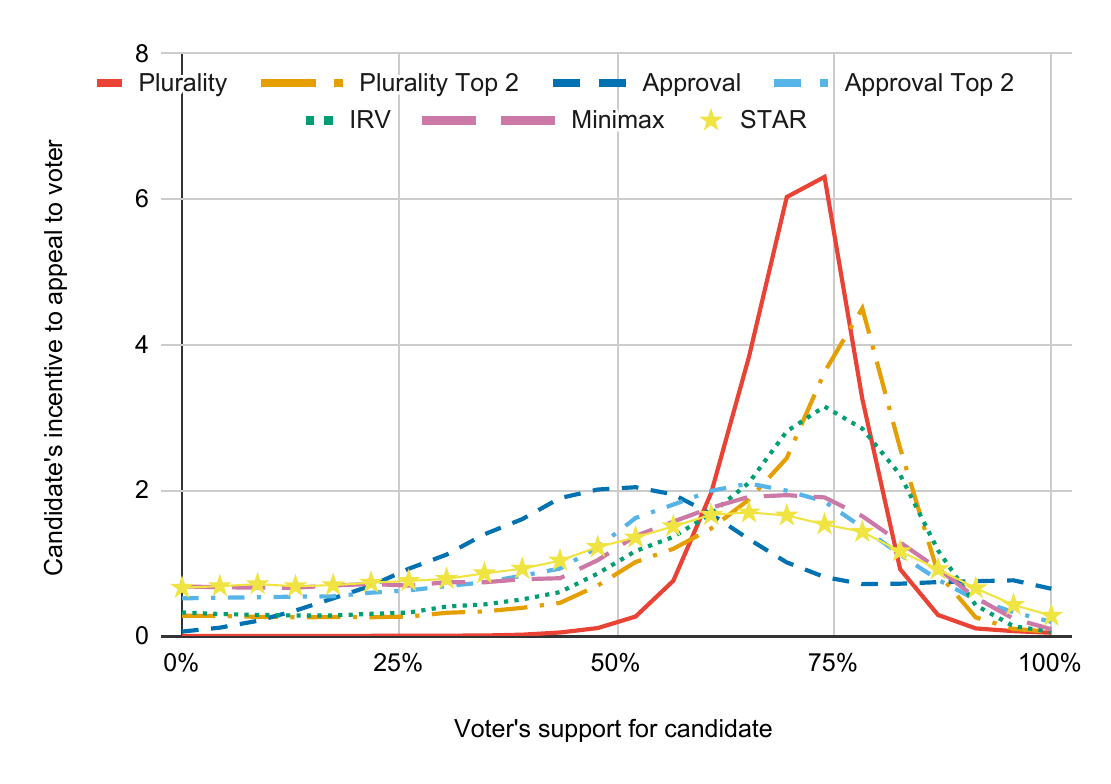}\caption{\label{fig:Distance-from-Top}Distance from Top, 5 candidates, sincere
voters, 50,000 iterations}
\end{figure}

In Figure \ref{fig:Distance-from-Top}, the peaks of Plurality, Plurality
Top 2, and IRV are much higher than with Normalized Mean. Distance
From Top is highly dependent on how voters feel about their first
choice, so it naturally leads to more vivid results with voting methods
that place heavy importance on being a voter's first choice. However,
this focus on first choices makes it worse for identifying exceptionally
influential voting blocs under voting methods that do not place special
emphasis on voters' first choices. If a voter strongly prefers a non-viable
candidate to all others, this has a major effect on Distance From
Top (the voter appears to be strongly opposed to all other candidates)
but a relatively minor effect on Normalized Mean. The presence of
the spoiler effect within the sorting function effectively adds noise
to the sorting function, which makes the CIDs of other voting methods
appear to be more even than they ought to. A ``sorting function''
that rearranged all voters at random would make all voting methods
appear to have a perfectly even CID, so we prefer to use sorting functions
that yield as little noise as possible.

\begin{figure}

\includegraphics[scale=0.7]{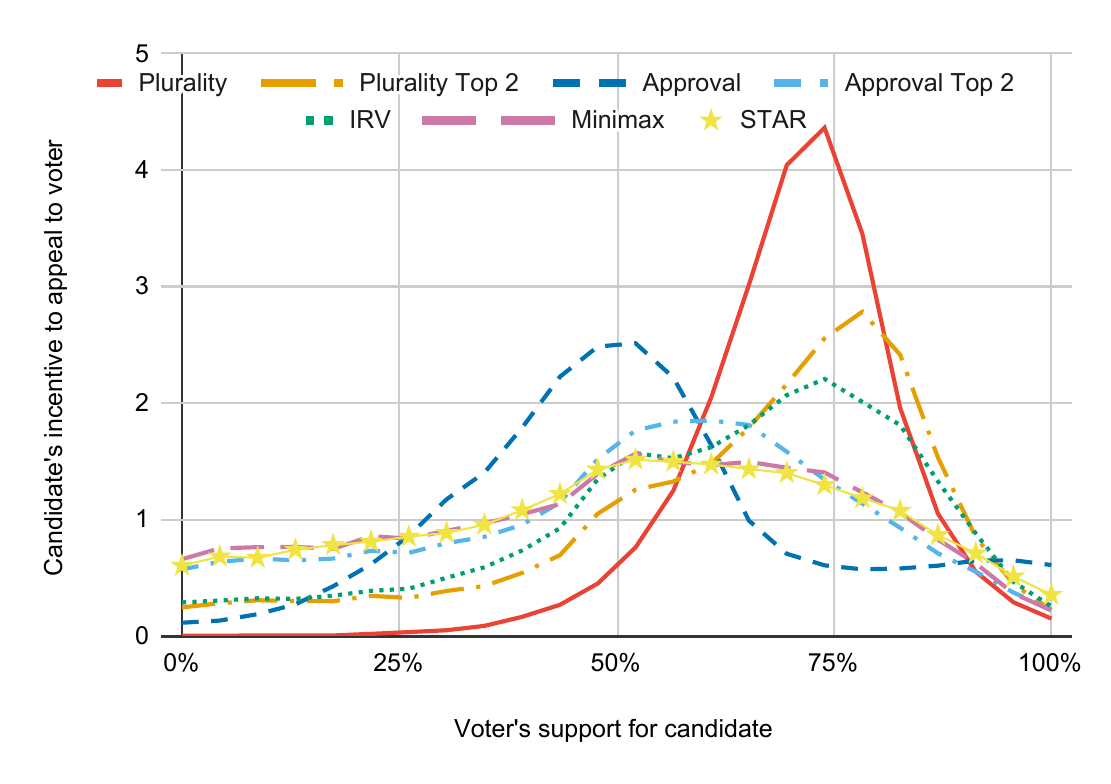}\caption{\label{fig:720-voters}720 sincere voters, 5 candidates, 25,000 iterations}

\end{figure}

Finally, we consider whether the number of voters makes a significant
difference. Since figures \ref{fig:720-voters} and \ref{fig:hon5c}
look nearly identical, we conclude that the choice to use only 72
voters in most simulations, rather than hundreds or thousands, has
an insignificant effect on the results.

\end{document}